\PassOptionsToPackage{table,dvipsnames}{xcolor}
\documentclass[11pt,a4paper,logo]{googledeepmind}

\setrightlogo[110pt]{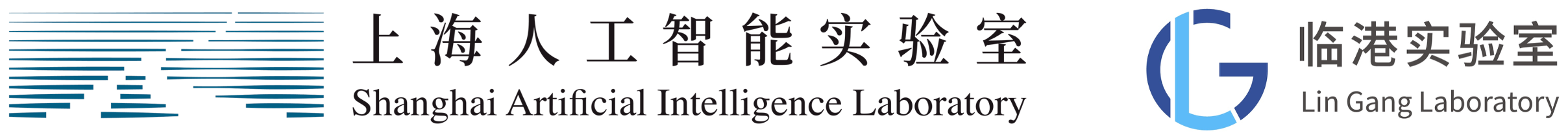}
\usepackage{fontawesome5}
\usepackage{marvosym}
\usepackage{academicons}
\usepackage{tgcursor}
\usepackage{pifont}
\usepackage[utf8]{inputenc} % allow utf-8 input
\usepackage[T1]{fontenc}    % use 8-bit T1 fonts
\usepackage{hyperref}       % hyperlinks
\usepackage{url}            % simple URL typesetting
\usepackage{booktabs}       % professional-quality tables
\usepackage{amsfonts}       % blackboard math symbols
\usepackage{nicefrac}       % compact symbols for 1/2, etc.
\usepackage{microtype}      % microtypography
\usepackage{graphicx}
\usepackage{amsmath}
\usepackage{tabularx}
\usepackage{multirow}
\usepackage{multicol}
\usepackage{wrapfig}
\usepackage{listings}
\usepackage{etoolbox}
\usepackage[
    natbib=true,
    backend=biber,
    style=numeric, % Changed from 'alphabetic' to 'numeric' as per common technical report styles
    sorting=none, % To keep references in order of appearance or as defined in .bib
    maxbibnames=5, % Show up to 5 authors before using et al
    minbibnames=5  % Show at least 5 authors before truncating
]{biblatex}
\AtEveryBibitem{\clearfield{month}}
\AtEveryBibitem{\clearfield{day}}
\usepackage{csquotes}

\title{EasyBCI Agent: Towards Universal Neural Data Preprocessing for Brain-Computer Interfaces}
\newcommand{\github}{\raisebox{-1.5pt}{\includegraphics[height=1em]{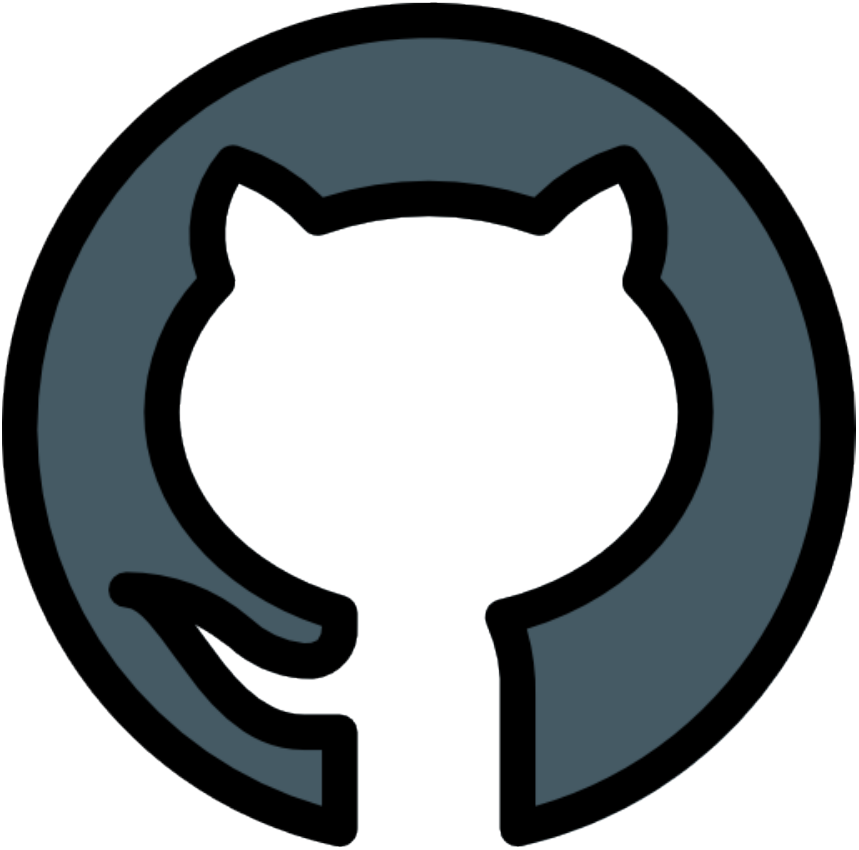}}}

\usepackage{pdflscape}

\usepackage{textcomp}
\usepackage{rotating}

\usepackage{setspace}
\usepackage{subcaption}
\usepackage{caption}

\usepackage{array}
\usepackage{threeparttable}
\usepackage{bm}

\usepackage{siunitx}

\usepackage{enumitem}
\usepackage{float}
\usepackage{seqsplit}
\usepackage{framed}

\usepackage{tikz}

\usepackage[export]{adjustbox}

\usepackage{ragged2e}
\usepackage[most]{tcolorbox}

\usepackage{makecell}
\usepackage{tabularray}

\usepackage{longtable}
\newcolumntype{Y}{>{\centering\arraybackslash}X}   % centered, wrapping X
\newcolumntype{C}[1]{>{\centering\arraybackslash}p{#1}} % centered, fixed width

\lstdefinestyle{seedcontent}{
  basicstyle=\ttfamily\small\color{blue!45!black},
  breaklines=true,
  breakatwhitespace=false,
  columns=fullflexible,
  keepspaces=true,
  showstringspaces=false,
  frame=single,
  framerule=0.2pt,
  rulecolor=\color{black!15},
  backgroundcolor=\color{black!2},
  xleftmargin=0.5em,
  xrightmargin=0.5em
}
\lstdefinestyle{judgeprompt}{
  basicstyle=\ttfamily\scriptsize,
  breaklines=true,
  breakatwhitespace=true,
  breakautoindent=false,
  breakindent=0pt,
  columns=fullflexible,
  keepspaces=true,
  frame=single,
  xleftmargin=1em,
  xrightmargin=1em,
  showstringspaces=false
}

\definecolor{CaseGreen}{HTML}{2E7D32} % deep green
\definecolor{CaseOrange}{HTML}{F57C00} % orange bar
\definecolor{CaseGray}{HTML}{F6F7F8}   % light gray bg
\definecolor{CaseInk}{HTML}{212121}    % dark text
\definecolor{CaseWhite}{HTML}{F5F5DC}
\definecolor{DeepBlue}{HTML}{003366} % 深蓝
\definecolor{LightBlue}{HTML}{99CCFF} % 浅蓝
\definecolor{DeepPurple}{HTML}{673AB7} % 深紫色
\definecolor{MiddlePurple}{HTML}{9C7FD0}
\definecolor{LightPurple}{HTML}{D1C4E9} % 浅紫色
\definecolor{HotPink}{HTML}{FF69B4} % 鲜粉色
\definecolor{SoftPink}{HTML}{F8BBD0} % 浅粉色
\definecolor{Crimson}{HTML}{DC143C} % 深红偏紫
\definecolor{Teal}{HTML}{008080} % 水鸭青
\definecolor{Cyan}{HTML}{00BCD4} % 青色
\definecolor{SoftGray}{HTML}{EEEEEE}       % 柔和浅灰
\definecolor{LighterGray}{HTML}{FAFAFA} % 超浅灰

\newtcolorbox{stagebox}[1]{
  breakable, enhanced,
  colback=CaseGray, colframe=DeepBlue, coltitle=CaseWhite,
  title=\bfseries #1, fonttitle=\bfseries,
  left=1.2mm,right=1.2mm,top=1.2mm,bottom=1.2mm, boxrule=0.6pt
}

\newtcblisting{verbatimbox}{
  listing only, breakable, enhanced,
  colback=white, colframe=LightBlue, boxrule=0.5pt,
  left=1.2mm,right=1.2mm,top=1.2mm,bottom=1.2mm
}

\usepackage[symbol]{footmisc}

\newcommand{\good}[1]{\textcolor{teal!70!black}{#1}}

\hypersetup{
    colorlinks=true,
    linkcolor=blue, % Changed for better visibility of links
    citecolor=blue,  % Changed for better visibility of citations
    filecolor=black,
    urlcolor=blue    % Changed for better visibility of URLs
}

\usepackage{tocloft}
\renewcommand{\github}{\raisebox{-1.5pt}{\includegraphics[height=1em]{figures/github-logo.pdf}}}

\usepackage{arydshln}
\makeatletter
\patchcmd{\@tocline}
    {\hfil}
    {\leaders\hbox{\hfil}\hfil}
    {}{}
\makeatother

\begin{abstract}
Brain-computer interfaces translate neural activity into device commands that can restore, enhance or replace lost function, yet their performance hinges on preprocessing of raw signals. This transformation remains manual, expert-dependent and poorly reproducible across laboratories. Large language model agents can now automate scientific coding, but existing systems do not jointly provide the modality coverage, raw-data isolation, experience accumulation and domain-expert oversight that neural preprocessing requires. To address these, we introduce \textbf{EasyBCI}, a two-phase LLM agent that plans and executes preprocessing pipelines for six signal types. A Plan Agent profiles each recording into a text-only Data Fingerprint that never exposes raw data to the model, and selects a literature-grounded operator sequence. An Execution Agent then generates, runs and self-corrects code until quality criteria are met, while a quality-gated experience system retains validated strategies as reusable skills and automatically deprecates entries whose performance degrades. A domain expert intervenes at two decision gates (plan confirmation and repair exhaustion), retaining human judgement at the points where undetected error can invalidate downstream analyses. Systematic evaluation on EEG with a fixed linear classifier indicates that all eight EasyBCI backbones preserve more task-relevant linear separability than the manually designed pipeline on the 4-class task, with six of eight exceeding it on binary classification. Under controlled same-backbone comparison, EasyBCI outperforms general-purpose coding agents on both label schemes for four of eight backbones. EasyBCI extends to five additional modalities spanning nearly three orders of magnitude in sampling rate, producing complete and reproducible pipelines for each with recorded decision provenance. These results indicate that domain-specific orchestration can bring auditable and reproducible preprocessing within reach of laboratories lacking dedicated preprocessing expertise. More broadly, the architecture illustrates design principles that may apply to AI agents in other scientific domains where preprocessing decisions shape downstream conclusions.
\end{abstract}

\makeatletter
\newcommand{\makeboxedfrontmatter}{%
\begin{tcolorbox}[
    enhanced,
    colback=cyan!5!white,
    colframe=cyan!5!white,
    boxrule=0pt,
    arc=10pt,
    left=18pt,
    right=18pt,
    top=5pt,
    bottom=0pt,
    width=\textwidth
]%
{\color{black}\normalfont\bfseries\fontsize{17}{20}\selectfont\raggedright\hyphenpenalty=10000\exhyphenpenalty=10000 \@title\par}
\vspace{0.7em}
{\normalfont\bfseries\fontsize{10}{13}\selectfont\raggedright\hyphenpenalty=10000\exhyphenpenalty=10000
Yu~Zhu\textsuperscript{1\normalsize$*$$\clubsuit$}\enspace
Runkai~Zhao\textsuperscript{1\normalsize$*$}\enspace
Zhimin~Zhou\textsuperscript{1\normalsize}\enspace
Jinyu~Cai\textsuperscript{1\normalsize}\enspace
Zhouheng~Yao\textsuperscript{1\normalsize}\enspace
Chutian~Zhang\textsuperscript{1}\\
Peiyuan~Li\textsuperscript{2\normalsize}\enspace
Xiaoxing~Zhang\textsuperscript{2\normalsize}\enspace
Qihao~Zheng\textsuperscript{1}\enspace
Jiamin~Wu\textsuperscript{1}\enspace
Mianxin~Liu\textsuperscript{1}\enspace
Chi~Zhang\textsuperscript{1}\enspace\\
Bo~Zhang\textsuperscript{1\normalsize}\enspace
Bin~Min\textsuperscript{2\normalsize}\enspace
Lei~Bai\textsuperscript{2\normalsize}\enspace
Chengyu~Li\textsuperscript{2\normalsize\textdagger\Letter}\enspace
Jingpeng~Wu\textsuperscript{2\normalsize\textdagger\Letter}\enspace
Chunfeng~Song\textsuperscript{1\normalsize\textdagger\Letter}\par}
\vspace{0.35em}
{\normalfont\fontsize{10.5}{13}\selectfont
\textsuperscript{1} Shanghai Artificial Intelligence Laboratory\quad
\textsuperscript{2} Lingang Laboratory\par}
\vspace{0.25em}
{\normalfont\fontsize{10.5}{13}\selectfont
\noindent
{\large $\clubsuit$} Project lead.\quad {\large*} Equal contribution. \quad{\large \textdagger} Corresponding author.\par}
\vspace{0.25em}
{\normalfont\fontsize{10}{12.5}\selectfont
\noindent
\Letter \quad songchunfeng@pjlab.org.cn,\enspace jingpeng.wu@lglab.ac.cn,\enspace chengyu.li@lglab.ac.cn\par}
\vspace{1.2em}
{\normalfont\fontsize{11}{13.5}\selectfont \theabstract\par}
\vspace{1.1em}
{\normalfont\fontsize{10.5}{13}\selectfont
\github\ \textbf{Code} \href{https://github.com/zhuyu-cs/EasyBCIdata-agent}{\texttt{EasyBCI-data-agent}}
\\}
\end{tcolorbox}
}
\makeatother

\titleformat{\section}
{\normalfont\bfseries\fontsize{20}{24}\selectfont}
{\thesection}
{1em}
{#1}
[]
\titlespacing*{\section}{0pt}{2.2ex plus4pt minus2pt}{1.2ex plus2pt minus1pt}

\begin{document}

\thispagestyle{empty}
\noindent
\hfill\includegraphics[width=214.5pt]{figures/logo.png}
\vspace{-5pt}
\par\noindent\rule{\textwidth}{0.4pt}
\vspace{2pt}

\makeboxedfrontmatter

% ---- Table of Contents ----
\clearpage
\begingroup
\hypersetup{linkcolor=black}
\renewcommand{\contentsname}{\centering Contents}
\setcounter{tocdepth}{2}
\tableofcontents
\endgroup
\clearpage

\section{Introduction}

Brain-computer interfaces (BCI) translate neural activity into device commands, enabling systems that restore lost motor function, augment cognition or establish new communication channels. These systems divide into invasive and non-invasive categories according to sensor proximity to neural tissue. Invasive approaches record from electrodes implanted on or within the cortex through modalities such as electrocorticography, stereo-electroencephalography and unit-level spike arrays. Recent milestones in this category include speech decoding at 97.5\% word accuracy across a 125,000-word vocabulary~\cite{card2024accurate}, independent home use sustained over nearly two years~\cite{card2026long}, conversational-rate communication~\cite{willett2023high, metzger2023high} and restored walking after chronic tetraplegia~\cite{lorach2023walking}. Non-invasive approaches acquire signals through scalp-mounted or optical sensors without surgical implantation, spanning modalities such as EEG, MEG and fNIRS. These modalities advance affective computing, sleep staging and closed-loop neuromodulation in parallel. Together, six recording modalities now populate the field, each capturing complementary facets of neural dynamics through distinct biophysical mechanisms~\cite{he2008multimodal}. All these results share a common analytical prerequisite whose automation remains limited. Before any decoder operates, raw neural signals must be transformed through a sequence of preprocessing decisions that determines what task-relevant information is preserved and what is discarded. Domain specialists still perform this transformation through manual parameter selection. Standardisation and documentation remain limited, making preprocessing a bottleneck for scientific reproducibility and cross-laboratory deployment.

\begin{figure*}[!t]
\centering
\includegraphics[width=\textwidth]{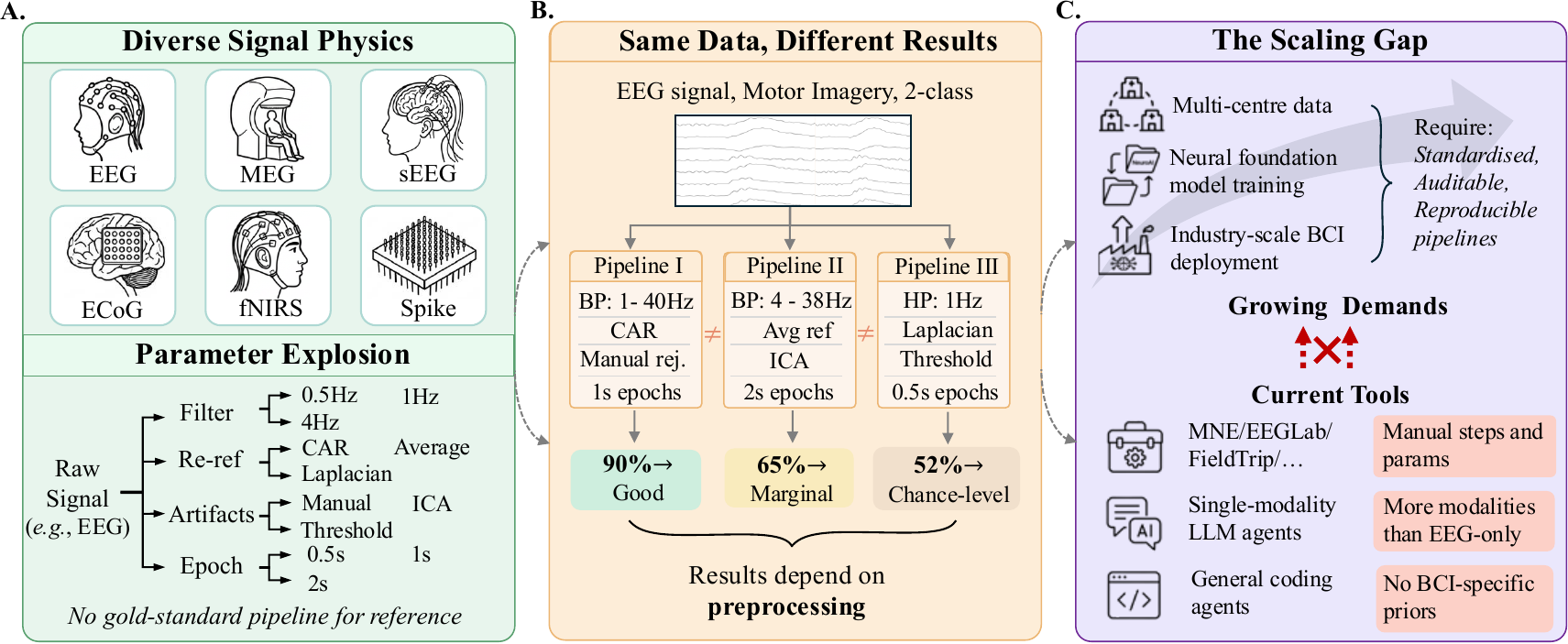}
\caption{\textbf{The preprocessing bottleneck in brain-computer interfaces.} (\textbf{A})~\textbf{Diverse Signal Physics.} Six recording modalities (EEG, MEG, sEEG, ECoG, fNIRS and unit-level spikes) span distinct biophysical mechanisms. Within each modality (e.g.\ EEG), preprocessing requires sequential choices over filtering, re-referencing, artifact rejection and epoching, creating a \textbf{parameter explosion} with no gold-standard pipeline for reference. (\textbf{B})~\textbf{Same Data, Different Results.} Three plausible pipelines applied to the same 2-class motor imagery recording with the same downstream classifier produce accuracies of 78\%, 65\% and 52\% (good, marginal and chance-level). Results depend on preprocessing. (\textbf{C})~\textbf{The Scaling Gap.} Growing demands from multi-centre data, neural foundation model training and industry-scale deployment require standardised, auditable and reproducible pipelines, yet current tools cannot bridge this gap. Signal-processing libraries rely on manual steps, single-modality LLM agents cover only EEG, and general coding agents lack BCI-specific priors.}
\label{fig:challenge}
\end{figure*}

This bottleneck arises because preprocessing admits multiple valid parameter choices at each step, and six modalities impose distinct processing logic (Fig.~\ref{fig:challenge}A). Volume-conduction correction in scalp EEG, haemodynamic denoising in fNIRS~\cite{pinti2019current, yucel2025fnirs} and hardware-specific spike sorting in intracortical arrays all follow different signal-processing principles. No single specialist covers this full range. The combinatorial parameter space within any one modality exceeds what documentation or intuition alone can manage, and no gold-standard pipeline exists for reference. The consequence is that the same data yield different results (Fig.~\ref{fig:challenge}B). Up to 42\% of trial-level predictions flip when only the preprocessing pipeline changes while the raw data and downstream classifier remain identical~\cite{hou2026same}. Independent preprocessing choices substantially shift decoding accuracy across paradigms~\cite{kessler2025eeg}, and independent laboratories routinely reach divergent conclusions from the same recordings. Preprocessing is not merely a preparatory step but an experimental variable whose parameter choices largely determine what information reaches the decoder and what conclusions a study can support.

BCI research now spans multi-centre data aggregation~\cite{liu2026open,seedat2025open}, large-scale foundation-model pretraining~\cite{jiang2024large,zhang2023brant}, clinical translation~\cite{bowsher2016brain,mitchell2023assessment} and long-term home use with unsupervised recalibration~\cite{card2026long,wilson2025long}. These developments demand standardised, auditable and reproducible pipelines that manage variation across data sources, centres and deployment settings (Fig.~\ref{fig:challenge}C). Neither existing toolboxes nor recent AI systems close this gap. Signal-processing libraries such as MNE-Python, EEGLAB and FieldTrip serve individual modalities well but provide no automated decision layer that selects among preprocessing options without expert intervention. LLM agents applied to EEG preprocessing~\cite{zhao2026eeg, abdou2026eeg} demonstrate that signal-processing orchestration is feasible~\cite{wang2024executable}, yet address only one modality. General-purpose coding agents such as Claude Code and Codex generate syntactically correct pipelines, but their parameter choices are not grounded in the signal characteristics of the recording at hand. Closing this gap requires a system with four capabilities. It must cover all six signal types. It must isolate raw data, because neural recordings carry personally identifiable physiological signatures~\cite{yuste2023advocating, zhong2025considerations}. It must retain validated strategies without degradation from naive accumulation~\cite{shi2026evolving}. It must embed expert oversight at defined decision points~\cite{takerngsaksiri2025human}. No existing system satisfies these four requirements jointly.

To address these challenges we introduce \textbf{EasyBCI}, a two-phase preprocessing agent driven by a large language model that plans and executes pipelines for all six modalities (Fig.~\ref{fig:framework}). A Plan Agent profiles each recording, encodes the result into a text-only Data Fingerprint that never exposes raw arrays to the model, and selects a literature-grounded operator sequence. An Execution Agent then generates, runs and self-corrects code through a Code, Execute and Reflect loop until quality criteria are met. Four architectural mechanisms support these phases. Modality-aware routing dispatches all six signal types through a single planner. A three-layer Source Data Guard isolates raw signals from the model. Quality-weighted Skill Retrieval selects the most relevant strategies from a growing Proven Skills library with quality-gated acceptance and automatic deprecation. An Expert-in-the-Agent-Loop protocol places domain-expert oversight at plan confirmation and repair exhaustion, retaining human judgement within otherwise autonomous execution.

To verify preprocessing quality, we adopt a fixed-classifier protocol. CSP features and LDA under five-fold cross-validation serve as a fixed-classifier probe of preserved task-relevant linear separability. Through systematic evaluation on EEG we compare EasyBCI instantiated with eight backbone LLMs against general-purpose coding agents and a manually designed pipeline. Under controlled same-backbone comparisons, the results indicate that domain-specific orchestration preserves more task-relevant linear separability than general-purpose agents given the same model. On the four-class task this advantage holds for five of eight backbones; on binary classification it holds for four of eight. We further apply EasyBCI to five additional modalities spanning nearly three orders of magnitude in sampling rate. This work makes the following contributions.

\begin{itemize}
\item A unified decision architecture that routes six neural signal modalities through automated, literature-grounded preprocessing, reducing the manual effort required for reproducible cross-modal studies.
\item A privacy-by-architecture principle in which raw arrays never enter the LLM context, so that privacy review reduces to inspecting a text fingerprint.
\item Quality-gated skill accumulation with automatic deprecation, enabling the knowledge base to improve empirically over successive runs.
\item An Expert-in-the-Agent-Loop protocol that retains domain-expert judgement at plan confirmation and repair exhaustion within otherwise autonomous execution.
\item A fixed-classifier verification methodology that isolates preprocessing as the sole experimental variable, demonstrated on EEG and applicable in principle to other modalities.
\end{itemize}

Beyond the six modalities tested here, the pluggable reader and skill design can in principle accommodate other neural signal types. More broadly, the design principles of modality-aware planning, raw-data isolation, experience accumulation and expert oversight at defined decision points may apply to other scientific workflows in which preprocessing decisions shape downstream conclusions.

\section{Related Work}
\label{sec:related}

EasyBCI combines LLM-based agent techniques with neurophysiological preprocessing knowledge.

\subsection{LLM-based Agents}
\label{sec:related-agents}

EasyBCI builds on two core agent capabilities, reasoning-and-acting loops and executable code generation. ReAct~\cite{yao2023react} introduced the pattern of interleaving chain-of-thought reasoning with tool calls. Reflexion~\cite{shinn2023reflexion} adds verbal self-critique to this loop, and CodeAct~\cite{wang2024executable} shows that executable Python is a more expressive action space than discrete tool-selection tokens. MetaGPT~\cite{hong2024metagpt} and AutoGen~\cite{wu2023autogen} organise multi-agent collaboration under role-based protocols. In software engineering, SWE-agent~\cite{yang2024swe} and OpenHands~\cite{wang2025openhands} demonstrate that agent-computer interfaces can resolve real GitHub issues, while self-debugging methods~\cite{chen2024teaching} introduce iterative repair. EasyBCI adopts the ReAct reasoning loop and CodeAct execution model. It specialises both to signal-processing operators with typed I/O contracts, introducing domain constraints that general-purpose settings lack.

Long-running agents must accumulate reusable experience without degradation. Voyager~\cite{wangvoyager} maintains a growing skill library validated by in-game task success. CLIN~\cite{majumder2023clin} retains natural-language insights across episodes. TroVE~\cite{wang2024trove} builds verifiable Python toolboxes validated by unit tests. Naive skill accumulation can degrade performance over time, as irrelevant or outdated entries dilute retrieval quality~\cite{shi2026evolving}. SkillBrew~\cite{hu2026skillbrew} addresses this by treating skill curation as a multi-objective optimisation that balances usefulness, diversity and coverage. SkillOps~\cite{pu2026skillops} further models skill libraries as self-maintaining ecosystems with typed contracts and lifecycle management including deprecation. EasyBCI adds a domain-specific quality gate tied to signal-level metrics. Only pipelines meeting neurophysiological criteria persist, and under-performing entries are automatically deprecated.

Rather than accumulating skills, some agents improve their own control logic. G{\"o}del Agent~\cite{yin2024g} lets an LLM recursively rewrite its own reasoning structure guided by high-level objectives. Self-Harness~\cite{zhang2026self} formalises this as an iterative loop of weakness mining, harness proposal and regression-validated acceptance. EasyBCI does not modify its own control logic; it retains experience as quality-gated skills whose matching weights and deprecation thresholds are fixed by design.

Human oversight within agent execution improves reliability. HULA~\cite{takerngsaksiri2025human} shows that an explicit human gate improves agent reliability on enterprise software tasks, and recent surveys map the broader agent landscape~\cite{xi2025rise,wang2024survey,ren2025towards,rao2026scidatasailor}. EasyBCI places the expert inside the agent loop at two defined decision gates, plan confirmation and repair exhaustion. This placement targets the two points where undetected error can propagate to downstream conclusions. Across this body of work, none addresses neurophysiological preprocessing.

\subsection{BCI-specific Agents}
\label{sec:related-bci-agents}

Several recent systems apply LLM capabilities to BCI data, each addressing a different segment of the analytical chain. EEG-GPT~\cite{kim2024eeg} explores chain-of-thought EEG classification but does not touch preprocessing. ChatBCI~\cite{hong2024chatbci} embeds an LLM into a P300 speller for language-assisted sentence composition. EEGUnity~\cite{qin2025eegunity} uses LLM tooling to harmonise dataset metadata rather than process raw signals. EEGAgent~\cite{zhao2026eeg} is among the first LLM agents that plan across EEG operators for perception, event detection and report generation. EEG-AI~\cite{abdou2026eeg} is the most closely related prior system. It pairs an LLM planner with expert feedback to run ICA-based artifact removal on EEG recordings.

EasyBCI differs from these systems on four axes. On modality coverage, EEGAgent and EEG-AI target EEG only, whereas EasyBCI routes EEG, MEG, sEEG, ECoG, fNIRS and unit-level spike recordings through a single planning architecture. On experience accumulation, prior BCI agents re-plan from scratch on each run, whereas EasyBCI saves each pipeline that passes quality control as a quality-weighted skill that guides future generation and automatically deprecates entries whose performance degrades. On raw-data isolation, prior BCI agents lack a documented data-protection mechanism, whereas EasyBCI passes only a text-only Data Fingerprint to the model and enforces a three-layer Source Data Guard that blocks writes to registered paths and verifies file integrity around each step. On domain-expert oversight, prior systems either lack a formal gate or place experts outside the execution loop, whereas EasyBCI supports explicit expert confirmation before plan execution and provides a second gate when automated repair is exhausted. These four differences form the domain-specific orchestration layer that EasyBCI adds atop the preprocessing toolbox ecosystem reviewed below.

\subsection{BCI and Electrophysiology Preprocessing Tools}
\label{sec:related-tools}

The operators that EasyBCI generates build on a mature set of open-source libraries. EEGLAB~\cite{delorme2004eeglab} introduced interactive ICA-based artifact removal for EEG. FieldTrip~\cite{oostenveld2011fieldtrip} and Brainstorm~\cite{tadel2011brainstorm,tadel2019meg} extended coverage to MEG, sEEG and ECoG within a unified source-imaging framework. MNE-Python~\cite{gramfort2013meg} is the Python-native reference for M/EEG and iEEG analysis. Neo~\cite{garcia2014neo} defines a format-agnostic object model, and SpikeInterface~\cite{buccino2020spikeinterface} unifies ten spike-sorting backends. NeuroKit2~\cite{makowski2021neurokit2} spans peripheral and central signals, while BrainFlow~\cite{brainflow_web} provides a device-agnostic streaming SDK.

A second generation of tools automates common preprocessing decisions within individual modalities. PREP~\cite{bigdely2015prep} standardises re-referencing and bad-channel detection. Autoreject~\cite{jas2017autoreject} learns per-sensor thresholds for trial rejection. HAPPE~\cite{gabard2018harvard} and HAPPILEE~\cite{lopez2022happilee} target high-artifact developmental EEG. RELAX~\cite{bailey2023introducing} automates cleaning for oscillatory analyses. osl-ephys~\cite{van2025osl} adds a configurable batch layer for M/EEG. MNE-NIRS~\cite{luke2021analysis} standardises fNIRS preprocessing, and iEEG-recon~\cite{lucas2024ieeg} handles electrode localisation. Data standards have extended BIDS to iEEG~\cite{holdgraf2019ieeg}, MEG~\cite{niso2018meg} and NIRS~\cite{luke2025nirs}, with MNE-BIDS~\cite{appelhoff2019mne} automating ingestion. The NWB ecosystem~\cite{rubel2022neurodata} provides a cross-modality data model, and Magland et al.~\cite{magland2025facilitating} recently demonstrated LLM-assisted analysis over NWB on DANDI.

These tools provide the operator foundations that EasyBCI orchestrates, but all still require a human expert to choose among their options for each recording. The automation they offer is format-specific or paradigm-specific, and none retrieves reusable validated recipes from prior runs. Magland et al.~\cite{magland2025facilitating} come closest to our design by connecting an LLM to a neurophysiology stack, but they target post-hoc exploratory analysis over curated NWB archives rather than upstream preprocessing decisions that largely determine what information reaches the decoder. EasyBCI closes this gap by coupling the agent mechanisms from Section~\ref{sec:related-agents} with the operator libraries reviewed here. It adds modality-aware planning, raw-data isolation, quality-gated experience accumulation and expert oversight at defined decision points to automate preprocessing across all six modalities.

\section{Method}

\begin{figure*}[t]
\centering
\includegraphics[width=\linewidth]{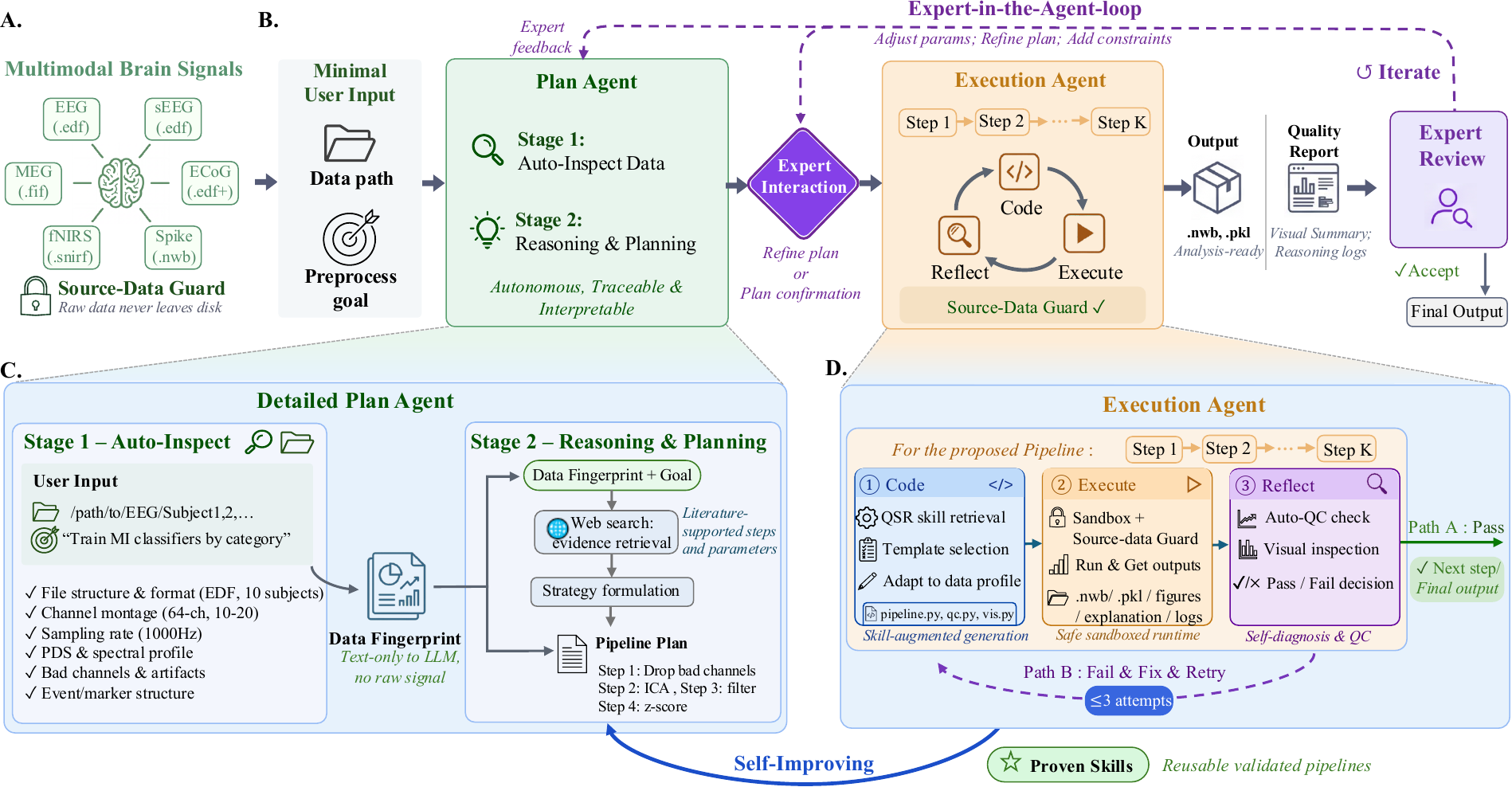}
\caption{\textbf{EasyBCI framework.} \textbf{A.} Multimodal brain signals (EEG, MEG, sEEG, ECoG, fNIRS and unit-level spikes) are ingested through a unified loader. The Source Data Guard ensures raw signal values never enter the LLM context. \textbf{B.} The user provides only a data path and a preprocessing goal in natural language. A Plan Agent extracts a Data Fingerprint and drafts a Pipeline Plan. An Expert Confirmation gate lets the domain expert confirm, modify or reject the plan before the Execution Agent carries it out step by step (Code, Execute, Reflect). The output (AI-ready NWB and pkl files, quality report and reasoning logs) is presented to the expert for final review. If the expert identifies issues, the outer loop iterates with adjusted parameters or constraints, forming a closed Expert-in-the-Agent-Loop. \textbf{C.} Detail of the Plan Agent. Stage 1 (Auto-Inspect) examines file structure, montage, sampling rate, spectral profile, artifact indicators and event structure to produce a text-only Data Fingerprint. Stage 2 (Reasoning and Planning) combines the fingerprint with web-retrieved literature evidence to emit a typed Pipeline Plan with per-step justifications. \textbf{D.} Detail of the Execution Agent. For each step in the approved pipeline, an inner loop runs three stages. (1) Code generates operator source guided by Quality-weighted Skill Retrieval. (2) Execute runs the code in a sandboxed subprocess under Source Data Guard protection. (3) Reflect applies automatic QC checks and visual inspection. On pass, the agent proceeds to the next step. On fail, it diagnoses the error and retries for at most three attempts before halting for expert intervention. Below the pipeline, Progressive Skill Refinement saves passing pipelines as Proven Skills and records failures as Negative Examples, with optional cross-lab Federation.}
\label{fig:framework}
\end{figure*}

EasyBCI converts raw neural recordings into AI-ready data through two phases (Fig.~\ref{fig:framework}). A \emph{Plan Agent} compresses the recording into a text-only Data Fingerprint and drafts a Pipeline Plan. An \emph{Execution Agent} carries out the approved plan step by step through a Code, Execute and Reflect loop. Four mechanisms support these phases, namely modality-aware routing, Source Data Guard, Quality-weighted Skill Retrieval and Expert-in-the-Agent-Loop.

The user provides only a data path and a natural-language preprocessing goal. When no goal is specified, the agent defaults to generic signal conditioning without epoch segmentation. The LLM never receives raw signal values and operates exclusively on the Data Fingerprint. Within the Execution Agent, an inner loop diagnoses and repairs failures for up to three attempts per step without human intervention. An outer loop lets the domain expert confirm or revise the plan before execution and review the final output, iterating only when automatic recovery is exhausted. This dual-loop structure reduces the expert's burden to two well-defined decision points. Plans draw on a growing library of Proven Skills that have passed quality control on similar data. The two phases share state through the local file system, which decouples plan approval from execution and allows resuming or editing a run at any point.

\subsection{Problem Formulation}
\label{sec:method-formulation}

Given a raw neural recording at path \(p\) and a natural-language preprocessing intent \(\tau\), the agent produces a preprocessing pipeline \(\pi = (o_1,\dots,o_K)\). The full output includes AI-ready arrays, a quality-control report and a self-contained mini-repository that can regenerate the result from \(p\). Formally, the core mapping is \((p,\tau) \mapsto \pi\). When \(\tau\) specifies a concrete downstream objective such as motor-imagery classification or speech decoding, the agent additionally produces labelled epoch arrays suitable for training a BCI decoder. When \(\tau\) contains no explicit objective, the agent restricts output to signal conditioning and derived features (e.g.\ firing-rate matrices) without AI-ready export.

\subsection{Operator Contract}
\label{sec:method-contract}

Composable pipelines require each operator to follow a uniform interface so that steps can be reordered, cached and validated independently. Every operator consumes and produces a state dict \(\mathcal{D}=(\mathbf{X},\,\mathbf{c},\,f_s,\,T,\,\mu)\), where \(\mathbf{X}\!\in\!\mathbb{R}^{C\times N}\) is a channel-by-time sample array, \(\mathbf{c}\) is the ordered channel-name list of length \(C\), \(f_s\) is the sampling frequency in hertz, \(T=N/f_s\) is the segment duration, and \(\mu\) carries per-step metadata (parameters, cache keys, elapsed time). Figure~\ref{fig:contract} shows the interface that all generated operators must implement.

\begin{figure}[!htbp]
\centering
\begin{tcolorbox}[enhanced, colback=CaseGray, colframe=DeepBlue!70, boxrule=0.4pt,
  left=2mm, right=2mm, top=1.5mm, bottom=1.5mm, width=\columnwidth,
  title={\small\bfseries Operator Contract (operator\_schema.py)}]
\ttfamily\scriptsize
\begin{tabular}{@{}l@{}}
\textcolor{DeepBlue}{@dataclass} \\
\textcolor{DeepBlue}{class} OperatorIO: \\
\quad data: np.ndarray \quad\textcolor{gray}{\# shape (C, N), float32/64} \\
\quad channels: List[str] \quad\textcolor{gray}{\# ordered names, len = C} \\
\quad frequency: float \quad\textcolor{gray}{\# sampling rate in Hz} \\
\quad duration: float \quad\textcolor{gray}{\# T = N / frequency} \\
\quad meta: Dict[str, Any] \quad\textcolor{gray}{\# per-step state (cumulative)} \\[4pt]
\textcolor{DeepBlue}{def} operator\_bandpass\_filter( \\
\quad\quad data\_dict: \textcolor{DeepPurple}{dict}, \\
\quad\quad *, \\
\quad\quad low: \textcolor{DeepPurple}{float} = 1.0, \\
\quad\quad high: \textcolor{DeepPurple}{float} = 40.0, \\
) -> \textcolor{DeepPurple}{dict}: \\[4pt]
\quad\textcolor{gray}{\# Rules enforced by AST linter (code\_standard\_check):} \\
\quad\textcolor{gray}{\# 1. Must not mutate data\_dict["data"] in place} \\
\quad\textcolor{gray}{\# 2. Must raise EasyBCIOperatorError on failure} \\
\quad\textcolor{gray}{\# 3. Must not perform network/file I/O} \\
\quad\textcolor{gray}{\# 4. Must use seeded RNG (EASYBCI\_SEED) if stochastic} \\
\quad ... \\
\quad \textcolor{DeepBlue}{return} new\_data\_dict \\
\end{tabular}
\end{tcolorbox}
\caption{\textbf{Operator contract.} Every generated operator is a plain function whose first positional argument is a state dict and whose return value is a new state dict. Immutability and I/O restrictions are enforced by a static AST linter at code-generation time.}
\label{fig:contract}
\end{figure}

Three rules govern all operators. First, an operator must not modify the input state in place. Second, it must raise a typed \texttt{EasyBCIOperatorError} on failure rather than returning silently corrupted output. Third, it must not perform network or file I/O unless explicitly marked with an allow directive. A static AST linter checks these rules at generation time and rejects any code that violates the contract before it can execute.

\subsection{Auto-Inspection and Data Fingerprint}
\label{sec:method-perception}

Raw neural signals are too large for the LLM context window and carry personally identifiable physiological content~\cite{yuste2023advocating}. The Plan Agent therefore compresses the recording into a short text summary that captures what the planner needs without revealing sensitive content (Fig.~\ref{fig:framework}C, Stage 1).

Let \(x\) denote the signal array loaded from \(p\). The loader dispatches on file extension and MNE channel type~\cite{gramfort2013meg} and computes a fingerprint
\begin{equation}
\phi(x) = \bigl\langle\, s,\,m,\,C,\,f_s,\,T,\,\Psi(x),\,\Gamma(x),\,\varepsilon(x),\,\eta(x)\,\bigr\rangle,
\label{eq:fingerprint}
\end{equation}
where \(s\) is the container format, \(m\) is the modality label, \(\Psi(x)\) summarises the spectral profile (power-line peak, harmonics, drift, high-frequency noise), \(\Gamma(x)\) summarises channel-level statistics (per-channel variance, flat-line ratio, spike count, drop candidates), \(\varepsilon(x)\) summarises artifact indicators (blink rate, muscle contamination percentage), and \(\eta(x)\) describes the event structure (event count, event types). Only \(\phi(x)\) enters the LLM context. The raw array \(\mathbf{X}\) remains in the local process and is never transmitted to the language model. Figure~\ref{fig:fingerprint} shows an abbreviated example of the JSON output for a 22-channel EEG recording.

\begin{figure}[!htbp]
\centering
\begin{tcolorbox}[enhanced, colback=CaseGray, colframe=DeepBlue!70, boxrule=0.4pt,
  left=2mm, right=2mm, top=1.5mm, bottom=1.5mm, width=\columnwidth,
  title={\small\bfseries InspectionReport (22-ch motor-imagery EEG, 250\,Hz)}]
\ttfamily\scriptsize
\begin{tabular}{@{}ll@{}}
\textcolor{DeepBlue}{"fingerprint":} \{ & \\
\quad "format" & \textcolor{DeepPurple}{"gdf"} \\
\quad "modality" & \textcolor{DeepPurple}{"eeg"} \\
\quad "n\_channels" & \textcolor{DeepPurple}{22} \\
\quad "sampling\_freq\_hz" & \textcolor{DeepPurple}{250.0} \\
\quad "duration\_s" & \textcolor{DeepPurple}{312.5} \\
\quad "n\_events" & \textcolor{DeepPurple}{288} \\
\quad "event\_types" & \textcolor{DeepPurple}{["769", "770", "771", "772"]} \\
\} & \\[3pt]
\textcolor{DeepBlue}{"psd\_summary":} \{ & \\
\quad "power\_line\_peak\_hz" & \textcolor{DeepPurple}{50.0} \\
\quad "power\_line\_peak\_db\_above\_floor" & \textcolor{DeepPurple}{8.7} \\
\quad "harmonics\_detected\_hz" & \textcolor{DeepPurple}{[100.0]} \\
\quad "low\_freq\_drift\_below\_1hz\_present" & \textcolor{DeepPurple}{true} \\
\quad "high\_freq\_noise\_above\_40hz\_present" & \textcolor{DeepPurple}{false} \\
\} & \\[3pt]
\textcolor{DeepBlue}{"channel\_summary":} \{ & \\
\quad "bad\_candidates\_high\_variance" & \textcolor{DeepPurple}{["Fz"]} \\
\quad "bad\_candidates\_flat" & \textcolor{DeepPurple}{[]} \\
\} & \\[3pt]
\textcolor{DeepBlue}{"artifact\_summary":} \{ & \\
\quad "blink\_rate\_per\_min" & \textcolor{DeepPurple}{14.2} \\
\quad "muscle\_artifact\_pct" & \textcolor{DeepPurple}{3.1} \\
\quad "saturation\_pct" & \textcolor{DeepPurple}{0.0} \\
\} & \\
\end{tabular}
\end{tcolorbox}
\caption{\textbf{Data Fingerprint.} Abbreviated JSON output of the Plan Agent's Auto-Inspect stage for a 22-channel motor-imagery EEG recording. Raw signal arrays are replaced by summary statistics, reducing token cost by three orders of magnitude while preserving all information needed for planning.}
\label{fig:fingerprint}
\end{figure}

\subsection{Reasoning and Planning}
\label{sec:method-planning}

Given \(\phi(x)\) and the user intent \(\tau\), the Plan Agent translates data characteristics and goals into a concrete operator sequence (Fig.~\ref{fig:framework}C, Stage 2). It outputs a Pipeline Plan
\begin{equation}
\mathrm{Plan} = \bigl\langle\, (o_k)_{k=1}^K,\,g,\,\theta,\,E,\,m,\,\kappa \,\bigr\rangle,
\label{eq:plan}
\end{equation}
where each \(o_k\) is a step string of the form \texttt{name} or \texttt{name:arg} drawn from a typed operator vocabulary, \(g\) is the analysis goal, \(\theta\) is a natural-language rationale, \(E\) is a list of literature evidence entries, \(m\) is the modality label, and \(\kappa\) is the paradigm tag (e.g.\ \texttt{motor\_imagery}, \texttt{auditory\_erp}). Each step string is validated against a registered operator vocabulary so that a plan can only reference operators with existing code implementations. Figure~\ref{fig:plan} shows an example plan generated for the motor-imagery EEG recording described in Figure~\ref{fig:fingerprint}.

\begin{figure}[!htbp]
\centering
\begin{tcolorbox}[enhanced, colback=CaseGray, colframe=DeepBlue!70, boxrule=0.4pt,
  left=2mm, right=2mm, top=1.5mm, bottom=1.5mm, width=\columnwidth,
  title={\small\bfseries PipelineProposal (motor-imagery EEG, BCI-IV-2a)}]
\ttfamily\scriptsize
\begin{tabular}{@{}ll@{}}
\textcolor{DeepBlue}{analysis\_goal} & \textcolor{DeepPurple}{"classification"} \\
\textcolor{DeepBlue}{modality} & \textcolor{DeepPurple}{"eeg"} \\
\textcolor{DeepBlue}{paradigm} & \textcolor{DeepPurple}{"motor\_imagery"} \\
\textcolor{DeepBlue}{steps} & \textcolor{gray}{\# compact operator:params grammar} \\
\quad 1. & \textcolor{DeepPurple}{"bandpass:4,38"} \\
\quad 2. & \textcolor{DeepPurple}{"ica:eog"} \\
\quad 3. & \textcolor{DeepPurple}{"drop\_bads:auto"} \\
\quad 4. & \textcolor{DeepPurple}{"epoch:0.5,2.5"} \\
\quad 5. & \textcolor{DeepPurple}{"export\_ai\_ready"} \\[3pt]
\textcolor{DeepBlue}{rationale} & \textcolor{gray}{Standard MI pipeline adapted to fingerprint.} \\
 & \textcolor{gray}{Moderate blink rate warrants ICA; low muscle} \\
 & \textcolor{gray}{contamination allows retaining up to 38\,Hz.} \\[3pt]
\textcolor{DeepBlue}{web\_evidence} & \\
\quad Ang et al., 2012 & \textcolor{gray}{4--38\,Hz bandpass for MI-BCI} \\
\quad Blankertz et al., 2008 & \textcolor{gray}{CSP optimal window 0.5--2.5\,s post-cue} \\
\end{tabular}
\end{tcolorbox}
\caption{\textbf{Pipeline Plan.} Example plan for a 22-channel motor-imagery EEG dataset. Steps use a compact string grammar (\texttt{operator:params}). The plan is presented to the domain expert at the confirmation gate before execution begins.}
\label{fig:plan}
\end{figure}

Evidence \(E\) comes from a web-search module that rewrites the fingerprint and goal into scholarly queries, retrieves results through a pluggable search backend, and produces per-step justifications with source citations. Literature evidence anchors parameter decisions to published experimental results, reducing the risk of unsupported parameter choices. The evidence is stored alongside the plan and shown to the expert at the confirmation gate (Section~\ref{sec:method-expert}).

\subsection{Code, Execute and Reflect}
\label{sec:method-execution}

Generated code may contain bugs, and data characteristics may violate assumptions made during planning. The Execution Agent therefore carries out the approved plan step by step, with each step \(o_k\) triggering a three-stage inner loop (Fig.~\ref{fig:framework}D).

In the \emph{Code} stage, the agent queries the Proven Skills library through Quality-weighted Skill Retrieval (Section~\ref{sec:method-qsr}). The matched skill metadata, including step list, parameter choices and success context, guides the LLM's code generation. The agent produces a Python function that conforms to the operator contract (Section~\ref{sec:method-contract}).

In the \emph{Execute} stage, the generated function is written to disk and run by a subprocess-isolated executor. The executor enforces an address-space limit (\texttt{RLIMIT\_AS}) and a wall-clock timeout, and the Source Data Guard (Section~\ref{sec:method-guard}) verifies source-file integrity around each call.

In the \emph{Reflect} stage, an automatic QC module examines the resulting \(\mathcal{D}\) and generated figures against a rule-based checklist covering channel variance range, spectrum sanity, event-alignment integrity and band-power drift. The module returns a pass or fail verdict for the current step.

When Reflect reports a failure, the error trace, the failing code segment and intermediate diagnostics are sent to an error classifier that routes to one of three recovery strategies. Rule-based pattern fixes handle known error signatures. Prior-error remedies draw from the failure archive (Section~\ref{sec:method-refinement}). Skill-based fixes draw on Proven Skills records. The selected strategy produces a code patch that is re-executed in the same sandboxed environment. The number of repair attempts per step is capped at \(r_{\max}=3\) with a persistent counter that survives restarts. If the budget is exhausted, the step halts and control returns to the expert (Section~\ref{sec:method-expert}). Platform-level problems (transient tool failures, provider timeouts) are retried on a separate counter, so infrastructure instability does not reduce the three attempts available for scientific errors.

\subsection{Quality-weighted Skill Retrieval}
\label{sec:method-qsr}

Generating preprocessing code from scratch for every recording introduces variance. Multiverse analyses show that pipeline choices alone can substantially shift EEG decoding accuracy~\cite{kessler2025eeg}. Reusing validated operator choices from prior runs on similar data reduces this variance (Fig.~\ref{fig:framework}D, Code stage). Quality-weighted Skill Retrieval is a rule-based experience-matching mechanism that uses no vector similarity or dense retrieval; matching operates over typed metadata fields with fixed dimension weights.

EasyBCI maintains a Proven Skills library \(\mathcal{S}=\{s_j\}_{j=1}^{|\mathcal{S}|}\), where each skill records modality, paradigm, channel count, sampling rate, duration, step list, analysis goal and an inline Python operator implementation. The library grows from two sources. The first is hand-authored operators shipped with EasyBCI. The second is pipelines automatically saved from prior successful runs (Section~\ref{sec:method-refinement}).

At code-generation time, EasyBCI finds the most relevant skills through a rule-based, dimension-weighted matching procedure. For a step \(o\) and a skill \(s_j\), a similarity score is computed as
\begin{equation}
\mathrm{sim}(o,\phi;s_j) = \sum_{d\in\mathcal{A}} w_d \cdot f_d\!\bigl(s_j^{(d)},\,(o,\phi)^{(d)}\bigr),
\label{eq:sim}
\end{equation}
where \(\mathcal{A}\) lists matching axes (modality, paradigm, channel count, sampling rate, duration, analysis goal), \(w_d\) are fixed dimension weights, \(s_j^{(d)}\) denotes the value of axis \(d\) in skill \(s_j\), and \(f_d \in [0,1]\) is a per-axis match function (exact equality for categorical axes, ratio-based proximity for numeric axes). Two axes, modality and analysis goal, act as hard filters that set the score to zero on mismatch.

The similarity is then scaled by a per-skill quality weight
\begin{equation}
\lambda(s_j) = \begin{cases}
0 & \text{if } s_j \text{ is deprecated,}\\
\lambda_{\min} & \text{if } s_j \text{ is manually flagged,}\\
\max\!\left(\lambda_{\min},\, \rho(s_j)/\rho^{*}\right) & \text{otherwise,}
\end{cases}
\label{eq:quality}
\end{equation}
where \(\rho(s_j)\) is the historical QC pass rate of \(s_j\) and \(\rho^{*}\) is a reference threshold. When federation is active, a per-lab diversity cap limits retrieval from any single source laboratory. A negative-example penalty from the failure archive further down-weights skills that share signatures with known bad plans. The top \(k=3\) skills by \(\mathrm{sim}(o,\phi;s_j)\!\cdot\!\lambda(s_j)\) are returned to the code-generation prompt as structured context (step lists, parameter choices and match rationale), which the LLM uses to guide code generation. Skills whose pass rate drops below 40\% after at least five uses are automatically deprecated.

\begin{figure}[t]
\centering
\includegraphics[width=\columnwidth]{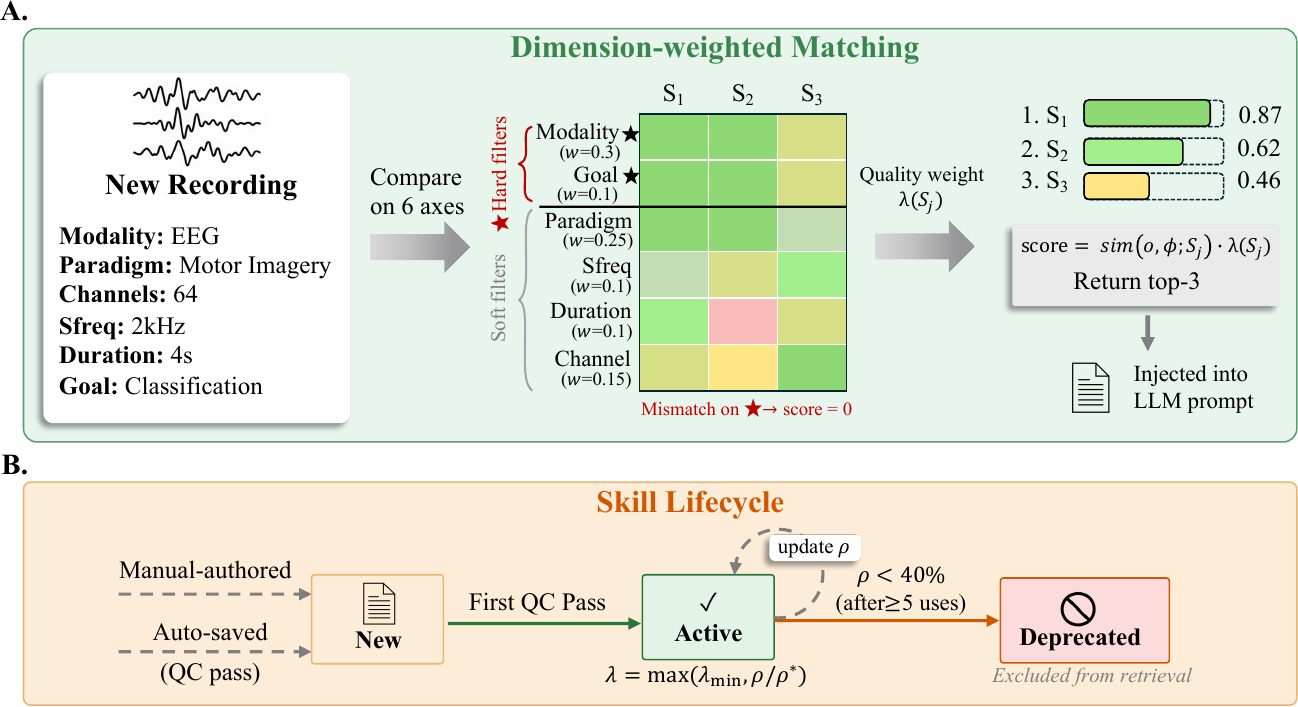}
\caption{\textbf{Quality-weighted Skill Retrieval and skill lifecycle.} (\textbf{A})~Dimension-weighted matching. A new recording's fingerprint is compared against library entries on six axes. Two hard filters (modality and analysis goal) reject mismatches outright. Four soft axes contribute weighted similarity scores visualised as a heatmap. The raw similarity is scaled by a per-skill quality weight $\lambda(s_j)$ and the top-$k$ results are injected into the LLM code-generation prompt as typed metadata. (\textbf{B})~Skill lifecycle. Skills enter the library through manual authoring or automatic saving after a QC pass. Active skills accumulate a historical pass rate $\rho$ that determines their quality weight. Skills whose pass rate drops below 40\% after at least five uses are deprecated and excluded from retrieval.}
\label{fig:skill-retrieval}
\end{figure}

The lifecycle of each skill follows a three-state progression from new to active to deprecated (Fig.~\ref{fig:skill-retrieval}A,B).

\subsection{Source Data Guard}
\label{sec:method-guard}

Raw brain recordings carry personally identifiable neural signatures~\cite{yuste2023advocating}, and any accidental modification would compromise reproducibility. EasyBCI enforces two invariants (Fig.~\ref{fig:framework}A,D). First, raw signal values never enter the LLM context. Second, raw files are never modified on disk. The Source Data Guard implements these invariants through three layers.

At the \emph{abstraction layer}, the LLM receives only \(\phi(x)\) and derived text. The numeric array \(\mathbf{X}\) is loaded in a Python subprocess and passed between operators as in-memory objects that never cross the language-model boundary.

At the \emph{write-block layer}, a registry of protected source paths is checked on every file-write operation. Any attempt to write, delete or rename inside a protected path raises a typed exception and aborts the current step.

At the \emph{integrity layer}, each source file's size and modification time are recorded before code execution and re-checked afterwards. A mismatch aborts the run and flags the step for investigation.

The executor runs generated code in a subprocess with a bounded address space (\texttt{RLIMIT\_AS}) and a wall-clock timeout, providing process-level isolation. A multi-tenant deployment would additionally require container-level sandboxing, which EasyBCI supports through a pluggable executor interface. For the single-researcher setting evaluated in this work, the three-layer guard is sufficient to keep raw data on disk and untouched throughout the pipeline.

\subsection{Expert-in-the-Agent-Loop}
\label{sec:method-expert}

Incorrect preprocessing choices can silently invalidate downstream analyses. EasyBCI embeds explicit expert intervention at the two highest-consequence decision points (Fig.~\ref{fig:framework}B, dashed loop).

The first is the \emph{plan confirmation gate}. After the Plan Agent produces a plan, EasyBCI presents it together with the fingerprint, literature evidence and matched skills. Execution blocks until the expert selects one of three actions, namely confirm, modify or abort. The expert's choice and any edits are logged for provenance.

The second is the \emph{repair exhaustion gate}. When a step's auto-repair budget is exhausted, the run pauses and presents the last error, failed patches and intermediate figures for expert diagnosis. The expert can supply a corrective hint, skip the step, or abort the run entirely.

At the end of each run, the QC module produces a human-readable report summarising per-step outcomes, and the expert can override the refinement loop's save decision.

\subsection{Progressive Skill Refinement and Federation}
\label{sec:method-refinement}

Without a retention mechanism, successful preprocessing decisions are lost after each run. EasyBCI turns each successful execution into a reusable Proven Skill (Fig.~\ref{fig:framework}D, bottom). A pipeline is saved to the library only when four conditions are jointly met. The QC grade must pass. The goal type must qualify for retention. The mini-repository must be contract-complete. No existing skill must already cover the same data signature. The saved skill carries the metadata listed in Section~\ref{sec:method-qsr}, and its future retrieval ranking is governed by the quality weight in Equation~(\ref{eq:quality}).

Failing runs also contribute to future retrieval. Each abort creates a negative example that records the plan signature and failure reason. At matching time, the negative archive adds a penalty to \(\lambda(s_j)\), pushing plans similar to known failures lower in the ranking. Per-use statistics are updated after each run, and skills whose pass rate drops below 40\% after five uses are automatically deprecated.

This retention-and-deprecation loop is motivated by recent findings that naive skill accumulation can degrade rather than improve agent performance~\cite{shi2026evolving}. Quality-gated acceptance and automatic deprecation mitigate this degradation. An alternative strategy is to let the agent rewrite its own control logic~\cite{zhang2026self, yin2024g}, but unconstrained self-modification risks violating the operator contract and data-protection invariants that neuroscientific preprocessing demands. Skill-level retention preserves these invariants while still allowing the agent to improve with use.

For cross-laboratory reuse, EasyBCI provides an opt-in federation module. Each participating lab holds an Ed25519 keypair. Proven skills can be signed and pushed to a shared git repository. Receiving labs pull, verify signatures under trust-on-first-use pinning, and merge verified skills into their local \(\mathcal{S}\). Skills with invalid signatures are quarantined rather than dropped. Federation is architecturally separate from the core execution loop and is not required for single-lab operation.

\subsection{Computational Reproducibility}
\label{sec:method-reproducibility}

Neural preprocessing pipelines frequently involve stochastic operations such as ICA decomposition, random train-test splits and bootstrap-based artifact detection. A single uncontrolled random state can produce different outputs on repeated runs, undermining scientific reproducibility. EasyBCI enforces deterministic execution through three mechanisms built into the code-generation and execution layers.

The code generator automatically inserts a deterministic seed preamble at the top of every generated script. This preamble locks three randomness sources simultaneously (the Python hash seed via an environment variable, the standard-library random module, and the NumPy global random state) to a fixed constant before any operator executes. The code-generation linter rejects any operator that calls stochastic functions without instantiating the RNG from this seed. Together these rules ensure that any pipeline produces identical outputs given identical inputs and the same software environment.

The same linter also forbids all generated scripts from importing any module under the EasyBCI framework namespace. The output mini-repository runs on a clean installation of five standard scientific Python packages (\texttt{mne}, \texttt{numpy}, \texttt{scipy}, \texttt{scikit-learn}, \texttt{matplotlib}) without requiring the agent to be installed. A blocking lint rule verifies this property at code-generation time. This decoupling ensures that any researcher can reproduce the same preprocessing by executing \texttt{python code/run.py} on the original recording without access to the agent, its configuration or its skill library.

Finally, the Source Data Guard's integrity layer (Section~\ref{sec:method-guard}) runs before and after every generated script, aborting on any change to source files. This prevents both accidental overwrite by buggy generated code and silent mutation by concurrent processes, ensuring that the input to a pipeline is the same file that was inspected during planning.

These three mechanisms operate without user configuration, producing output that is reproducible, self-contained and verifiably linked to the original source recording.

\subsection{Modality Coverage}
\label{sec:method-modality}

EasyBCI accommodates signal types that differ in sampling rate, channel semantics, artifact profiles and file-format conventions. A two-tier design separates dispatch logic from modality-specific knowledge (Fig.~\ref{fig:framework}A).

The first tier is a native loader registry. Each entry maps a file-format identifier and an MNE channel type to a reader function and a set of modality-aware defaults including expected frequency bands, reference conventions and artifact indicators. The registry currently covers EEG, MEG, sEEG, ECoG and unit-level spike recordings. Adding a new modality at this tier requires only a conforming reader that returns the standard state dict \(\mathcal{D}\) (Section~\ref{sec:method-contract}) and a metadata descriptor specifying sampling-rate range, channel semantics and applicable operator vocabulary.

The second tier is a skill-based interface that extends coverage without modifying core code. A modality skill file declares the signal type, the reader invocation (e.g.\ \texttt{mne\_nirs.io.read\_raw\_snirf} for fNIRS), format-specific preprocessing constraints and representative operator sequences. At planning time, the Plan Agent retrieves the relevant skill through the same matching mechanism used for Proven Skills (Section~\ref{sec:method-qsr}) and uses it to guide code generation. This tier currently supports fNIRS and vendor-specific spike-acquisition formats from Blackrock, Neuralynx, Plexon, SpikeGLX and OpenEphys.

\begin{table}[t]
\centering
\caption{\textbf{Six-modality coverage summary.} For each supported modality, the table lists typical sampling rate range, representative channel count, primary artifact sources, key preprocessing operations and the integration tier in EasyBCI.}
\label{tab:modality}
\small
\begin{tabular}{@{}llllll@{}}
\toprule
Modality & Sfreq (Hz) & Channels & Primary artifacts & Key operations & Tier \\
\midrule
EEG & 250--2048 & 22--256 & Blinks, muscle, line noise & Filter, ICA, re-ref, epoch & Native \\
MEG & 600--2400 & 102--306 & Heartbeat, eye, HPI & tSSS, ICA, source proj & Native \\
sEEG & 512--2048 & 8--256 & Line noise, HFO overlap & Bipolar re-ref, notch, epoch & Native \\
ECoG & 512--30000 & 16--128 & Line noise, electrode drift & CAR, notch, HG extraction & Native \\
Spike & 20000--40000 & 32--384 & Noise clusters, drift & Detection, sorting, rate est. & Native \\
fNIRS & 5--50 & 20--120 & Motion, systemic physiol. & TDDR, BPF, OD$\to$HbX & Skill \\
\bottomrule
\end{tabular}
\end{table}

Table~\ref{tab:modality} summarises the six modalities currently supported, their characteristic signal properties and the preprocessing operations that the planner draws from for each. A novel modality can be integrated by providing a conforming skill descriptor and a reader function that maps the raw file to \(\mathcal{D}\). No change to the planning logic, execution loop, quality-control rules or skill-retrieval mechanism is required. Modalities that mature from experimental to routine use can be promoted from skill-based to native loading without altering the agent's external behaviour.
\section{Experiments}
\label{sec:experiments}

We evaluate EasyBCI through a quantitative comparison on EEG motor imagery data and qualitative case studies across five additional modalities. We fix both the feature extraction method and the classifier across all conditions so that the preprocessing pipeline is the sole variable influencing downstream linear separability.

\subsection{Experimental Setup}
\label{sec:setup}

\subsubsection{Dataset}
\label{sec:datasets}

\begin{wraptable}{r}{0.48\columnwidth}
\vspace{-10pt}
\centering
\caption{\textbf{Primary EEG dataset.}}
\label{tab:dataset-eeg}
\footnotesize
\begin{tabular}{@{}l l@{}}
\toprule
\textbf{Property} & \textbf{Value} \\
\midrule
Modality & Scalp EEG \\
Paradigm & Object-directed motor imagery \\
Channels & 64 EEG + 2 EOG + 1 Trigger \\
Sampling rate & 1000~Hz \\
Sessions & 2 (37.6~min and 24.3~min) \\
Trials per session & 40 \\
Action type & 2-class (pick, place) \\
Object type & 4-class (apple, basket, bowl, plate) \\
Format & Curry (.cdt) \\
\bottomrule
\end{tabular}
\vspace{-8pt}
\end{wraptable}

The primary quantitative evaluation uses a 64-channel motor imagery EEG dataset acquired in our laboratory (Table~\ref{tab:dataset-eeg}). The recording consists of two sessions in Curry format, sampled at 1000~Hz with bipolar VEOG and HEOG references and one digital Trigger channel. Each session contains 40 cued motor imagery trials labelled by action type (pick and place, two classes) and object type (apple, basket, bowl and plate, four classes). This dataset exercises a representative preprocessing pipeline (filtering, artifact rejection, ICA, re-referencing, resampling and epoching). Existing BCI agent baselines support only EEG, making scalp EEG the natural choice for quantitative comparison.

Five additional modalities serve as qualitative case studies that evaluate cross-modal generalisation (Table~\ref{tab:case-summary}). Evaluation for these cases relies on signal-level QC metrics and domain-appropriate diagnostics rather than downstream classification accuracy.

\subsubsection{Evaluation Protocol}
\label{sec:eval-protocol}

The evaluation isolates preprocessing as the sole experimental variable through three design choices.

\paragraph{Fixed feature extraction.}
We apply Common Spatial Patterns (CSP) with six filter pairs followed by log-variance computation, yielding a 12-dimensional feature vector per trial. This feature pipeline is identical across all preprocessing conditions.

\paragraph{Fixed classifier.}
We use Fisher Linear Discriminant Analysis (LDA, implemented with solver \texttt{svd}) for all evaluations\footnote{\texttt{sklearn.discriminant\_analysis.LinearDiscriminantAnalysis}}. LDA has minimal model capacity and cannot compensate for poor preprocessing through nonlinear fitting, making it a fixed-classifier probe rather than a classification objective. Its accuracy measures the task-relevant linear separability that preprocessing preserves in the feature space.

\paragraph{Cross-validation and statistics.}
We perform stratified 5-fold cross-validation within each (session, label-scheme) combination, and repeat with three random split seeds for stability. The two label schemes are evaluated independently, action type (binary, two motor imagery classes) and object type (four classes). The reported accuracy for each condition and label scheme is the mean over the $2\times3=6$ per-condition CV-mean accuracies (2 sessions $\times$ 3 seeds).

\subsubsection{Compared Systems}
\label{sec:baselines}

We compare three categories of systems (Table~\ref{tab:main-results}). The first category contains two reference conditions. Raw applies no preprocessing and establishes a lower performance bound. Manual uses a conventional preprocessing pipeline applied uniformly to both sessions.\footnote{Designed by a doctoral researcher following published motor imagery EEG recommendations. The pipeline was fixed before execution without adaptation to the specific recording.} The second category consists of general-purpose coding agents that pair a large language model with a code-generation agent framework but lack domain-specific preprocessing knowledge. Claude Code pairs Claude Opus 4.8 with a general-purpose code-generation and execution loop. Codex pairs GPT-5.5 with OpenAI's autonomous software engineering agent. Both receive the same natural-language prompt describing the data and the preprocessing goal. The third category is EasyBCI itself, evaluated with eight backbone LLMs (Gemma4~\cite{team2026gemma}, Qwen3.6, Agents-A1~\cite{bai2026scaling}, GPT-4o, Opus 4.8, GPT-5.5, GPT-5.5-Pro and DeepSeek-V4-Pro~\cite{xu2026deepseek}) in cold-start mode with an empty skill library. This isolates the contribution of the planner, Code, Execute and Reflect loop and Source Data Guard from any benefit of an accumulated Proven Skills library.

This design tests two claims. First, domain-specific orchestration (iterative repair, literature grounding, modality-aware routing) preserves more task-relevant linear separability than general-purpose coding agents given the same backbone LLM. Second, among frontier-scale backbones, EasyBCI delivers stable quality across different backbone LLMs, so that agent architecture accounts for more variance than the choice of language model.

\subsubsection{Implementation Details}
\label{sec:implementation}

EasyBCI uses the same backbone LLM for both the Plan Agent and the Execution Agent within each run. For the main comparison we evaluate eight backbone LLMs (Gemma4, Qwen3.6, Agents-A1, GPT-4o, Opus 4.8, GPT-5.5, GPT-5.5-Pro and DeepSeek-V4-Pro) to separate architecture contribution from model capability. Among these, Gemma4 is a 31B dense model, Qwen3.6 and Agents-A1 are 35B mixture-of-experts models with 3B active parameters, and the remaining five are frontier-scale models. Claude Code uses Claude Opus 4.8 with its default agent loop and tool set. Codex uses GPT-5.5 in its default autonomous mode. All systems receive identical natural-language prompts describing the data and the preprocessing goal, and all have web search enabled. This ensures the comparison isolates agent architecture rather than information access. All experiments run on a single workstation (Intel Xeon, 64~GB RAM) without GPU acceleration. All EasyBCI runs start with an empty skill library. Web evidence retrieval uses Exa as the pluggable search backend for EasyBCI (Section~\ref{sec:method-planning}). Each pipeline step allows at most three automatic repair attempts before halting for expert intervention. The three sub-40B backbones each require multiple runs to produce a successful pipeline; all frontier-scale backbones succeed in a single run.

\subsection{EEG Preprocessing Results}
\label{sec:eeg-results}

\subsubsection{Downstream Classification Accuracy}
\label{sec:quantitative}

Table~\ref{tab:main-results} reports the linear separability measured by the fixed-classifier probe under two independent label schemes, action type (binary) and object type (4-class).

\begin{table}[t]
\centering
\caption{\textbf{Experimental conditions and preprocessing quality (LDA accuracy, \%).} All conditions share the same fixed feature extraction (CSP, 6 pairs, log-variance) and LDA classifier, isolating the effect of preprocessing from the underlying LLM. Mean $\pm$ std over 2 sessions $\times$ 3 random split seeds, each with stratified 5-fold cross-validation. Best in \textbf{bold}, second best \underline{underlined}.}
\label{tab:main-results}
\footnotesize
\begin{tabular}{@{}l l l c c@{}}
\toprule
\textbf{Condition} & \textbf{LLM} & \textbf{Preprocessing} & \textbf{Action (Binary)} & \textbf{Object (4-class)} \\
\midrule
Raw & -- & None & 49.95 $\pm$ 6.65 & 23.75 $\pm$ 9.87 \\
Manual & -- & Conventional & 55.25 $\pm$ 8.04 & 25.25 $\pm$ 10.18 \\
\midrule
Claude Code & Opus 4.8 & General agent & 64.00 $\pm$ 5.59 & 32.50 $\pm$ 11.68 \\
Codex & GPT-5.5 & General agent & 63.29 $\pm$ 2.82 & 35.54 $\pm$ 9.43 \\
\midrule
\textbf{EasyBCI} & Gemma4-31B-it~\cite{team2026gemma} & BCI agent & 50.00 $\pm$ 11.18 & 26.25 $\pm$ 10.75 \\
\textbf{EasyBCI} & Qwen3.6 35B-A3B & BCI agent & 48.75 $\pm$ 7.29 & 30.00 $\pm$ 10.75 \\
\textbf{EasyBCI} & Agents-A1-35B-A3B~\cite{bai2026scaling} & BCI agent & \textbf{56.25 $\pm$ 12.50} & \textbf{33.75 $\pm$ 7.50} \\
\midrule
\textbf{EasyBCI} & GPT-4o & BCI agent & 61.75 $\pm$ 8.09 & 36.65 $\pm$ 6.07 \\
\textbf{EasyBCI} & GPT-5.5 & BCI agent & 65.35 $\pm$ 2.07 & 40.65 $\pm$ 4.85 \\
\textbf{EasyBCI} & GPT-5.5-Pro & BCI agent & \underline{70.25} $\pm$ 2.07 & 43.55 $\pm$ 5.64 \\
\textbf{EasyBCI} & Opus 4.8 & BCI agent & 66.85 $\pm$ 4.72 & \underline{44.70} $\pm$ 5.60 \\
\textbf{EasyBCI} & DeepSeek-V4-Pro~\cite{xu2026deepseek} & BCI agent & \textbf{78.25} $\pm$ 4.90 & \textbf{46.65} $\pm$ 8.19 \\
\bottomrule
\end{tabular}
\end{table}

On the 4-class task all eight EasyBCI backbones outperform the Manual pipeline; on binary classification six of eight do so, with Gemma4 and Qwen3.6 falling below. Five of eight backbones exceed both general-purpose coding agents on the 4-class task, while four of eight do so on binary. DeepSeek-V4-Pro~\cite{xu2026deepseek} reaches 78.25\% on binary, 23 points above the Manual pipeline, and 46.65\% on the 4-class task versus 25.25\% for Manual. The three sub-40B backbones occupy the bottom three positions on both tasks, with Agents-A1~\cite{bai2026scaling} highest among them (56.25\% binary, 33.75\% 4-class). The middle ranks of Opus 4.8 and GPT-5.5-Pro swap between label schemes. Raw data yields near-chance accuracy (49.95\% binary, 23.75\% 4-class), indicating that the task requires effective preprocessing.

Under controlled backbone comparison, EasyBCI with Opus 4.8 scores 66.85\% versus 64.00\% for Claude Code (same LLM), and EasyBCI with GPT-5.5 scores 65.35\% versus 63.29\% for Codex (same LLM). The gap widens on the 4-class task (44.70\% vs.\ 32.50\% and 40.65\% vs.\ 35.54\%). On this dataset, domain-specific orchestration yields higher linear separability than general-purpose coding agents given the same underlying model. All five frontier-scale backbones exceed both general-purpose agents on the 4-class task, and four of five do so on binary. The three sub-40B backbones rank below both general-purpose agents on binary classification. Among them, Agents-A1~\cite{bai2026scaling} scores highest on both tasks and exceeds Claude Code on the 4-class task (33.75\% vs.\ 32.50\%).

\subsubsection{Qualitative Pipeline Demonstration}
\label{sec:qualitative}

\begin{table}[t]
\centering
\caption{\textbf{Pipeline comparison: Manual vs.\ EasyBCI (DeepSeek-V4-Pro).} The Manual pipeline applies five conventional steps and retains all 64 EEG channels. EasyBCI adds bad-channel removal, common average reference, ICA artifact rejection and robust scaling, producing a nine-step pipeline that removes five noisy frontal channels, suppresses ocular and cardiac artifacts via ICA and drops non-EEG channels, yielding 59 channels for feature extraction.}
\label{tab:expert-pipeline}
\scriptsize
\setlength{\tabcolsep}{3pt}
\begin{tabular}{@{}c l l c l l@{}}
\toprule
\multicolumn{3}{c}{\textbf{Manual (5 steps)}} & \multicolumn{3}{c}{\textbf{EasyBCI (9 steps)}} \\
\cmidrule(lr){1-3} \cmidrule(lr){4-6}
\textbf{\#} & \textbf{Operator} & \textbf{Params} & \textbf{\#} & \textbf{Operator} & \textbf{Params} \\
\midrule
1 & \texttt{drop\_misc} & VEOG, HEOG, Trig & 1 & \texttt{drop\_nondata} & markers only \\
2 & \texttt{notch} & 50~Hz & 2 & \texttt{notch} & 50~Hz \\
3 & \texttt{bandpass} & 0.4--40~Hz & 3 & \texttt{bandpass} & 0.5--45~Hz \\
4 & \texttt{resample} & 200~Hz & 4 & \texttt{drop\_bads} & auto \\
5 & \texttt{epoch} & 4.0~s & 5 & \texttt{car} & -- \\
  &  &  & 6 & \texttt{ica} & eog, ecg \\
  &  &  & 7 & \texttt{drop\_nondata} & data only \\
  &  &  & 8 & \texttt{resample} & 256~Hz \\
  &  &  & 9 & \texttt{scale} & robust \\
\bottomrule
\end{tabular}
\end{table}

Table~\ref{tab:expert-pipeline} compares the Manual pipeline with the best-performing EasyBCI pipeline (DeepSeek-V4-Pro backbone) on the motor imagery dataset. The Manual pipeline applies five steps and retains all 64 EEG channels without artifact rejection or re-referencing. EasyBCI introduces four additional operations (bad-channel removal, common average reference, ICA-based ocular and cardiac artifact rejection, and robust per-channel scaling). These additions correspond to the 23-point accuracy gap on the binary task (78.25\% vs.\ 55.25\%). The gap correlates with pipeline completeness. The Manual pipeline does not include operations that respond to recording-specific conditions such as the elevated frontal-channel variance and high blink rate in this dataset. EasyBCI derives each additional step from signal evidence surfaced during Auto-Inspect, reducing dependence on prior paradigm-specific experience. Epoch segmentation is handled identically across all conditions in a separate AI-ready export stage (Fig.~\ref{fig:code-structure}) and is not counted as a preprocessing step.

\paragraph{User instruction.}
The user issues a single natural-language instruction to the agent.

\begin{quote}
\small\itshape
Preprocess the 2-session EEG data in \texttt{./EEG/} for classification. Each session will be used to train two separate classifiers, one by action type and one by object type. 
\end{quote}

\paragraph{Evidence-grounded planning.}
The Plan Agent compresses the recording into a text-only Data Fingerprint (Section~\ref{sec:method-perception}). The fingerprint identifies 67~channels (64 EEG, 2 EOG, 1 trigger) and a 50~Hz power-line peak at 45.8~dB above the noise floor with harmonics at 100, 150 and 200~Hz. Session~2 shows 1101 blink-like detector events per minute (the threshold-based heuristic may overcount relative to true physiological blink rate). Five channels exceed 3 to 8 times the montage median variance (Fp1, Fpz, Fp2, AF3, AF4). From this fingerprint and the stated goal, the Reasoning and Planning stage issues a scholarly search through the pluggable web-evidence backend, retrieves citations from four sources, and emits a nine-step Pipeline Plan (Table~\ref{tab:plan}). Each step carries a confidence score that quantifies the strength of evidence for the chosen parameter value.

The plan derives parameters from signal evidence in three ways. First, values come from inspected signal characteristics. The notch frequency is set to 50~Hz because the power spectral density reveals a dominant peak at that frequency. Second, step ordering encodes causal dependencies. The notch precedes the bandpass to prevent side-lobe leakage into the analysis band. The Common Average Reference precedes ICA because source separation is more interpretable under a neutral reference. Resampling follows ICA to preserve the temporal resolution that FastICA convergence requires. Third, where no algorithmic derivation fully determines a parameter (e.g.\ the bandpass bounds 0.5 to 45~Hz), the web-evidence module provides literature justification with source URLs, making the choice externally auditable.

\begin{table*}[!htbp]
\centering
\caption{\textbf{Agent-generated pipeline plan for 64-channel motor imagery EEG.} Each row lists the operator, parameters, abridged rationale, evidence source and confidence score. The plan is produced without human input. The expert confirmation gate (Section~\ref{sec:method-expert}) serves as the single downstream checkpoint.}
\label{tab:plan}
\footnotesize
\setlength{\tabcolsep}{4pt}
\begin{tabularx}{\textwidth}{@{}c l l X l c@{}}
\toprule
\textbf{\#} & \textbf{Operator} & \textbf{Parameters} & \textbf{Rationale (abridged)} & \textbf{Evidence} & \textbf{Conf.} \\
\midrule
1 & \texttt{drop\_nondata} & mode=markers & Trigger is 99.96\% flat and removing it prevents CAR contamination & Inspection & 0.95 \\
2 & \texttt{notch} & freq=50 & 50~Hz peak at 45.8~dB above noise floor on the China 50~Hz grid & Inspection + Web & 0.95 \\
3 & \texttt{bandpass} & 0.5 to 45~Hz & Removes drift and EMG while retaining $\delta$ through low-$\gamma$ bands & Web\textsuperscript{\dag} & 0.85 \\
4 & \texttt{drop\_bads} & mode=auto & Fp1, Fpz, Fp2, AF3, AF4 exceed $3\times$ median variance & Inspection & 0.90 \\
5 & \texttt{car} & & 59 EEG channels remain after removing 5 bad electrodes, well above the $\geq$16-channel stability threshold for CAR & Domain skill & 0.90 \\
6 & \texttt{ica} & eog, ecg & 1101 blink-like detector events per minute; VEOG/HEOG serve as reference signals & Inspection + Web & 0.90 \\
7 & \texttt{drop\_nondata} & mode=data\_only & EOG channels are redundant after ICA artifact removal & Domain skill & 0.95 \\
8 & \texttt{resample} & 256~Hz & Nyquist-safe for 45~Hz upper bound with $4\times$ size reduction & Domain skill & 0.90 \\
9 & \texttt{scale} & robust & Per-channel median and IQR normalisation that is outlier-resistant & Domain skill & 0.85 \\
\bottomrule
\end{tabularx}
\vspace{2pt}
\hfill{\scriptsize \textsuperscript{\dag}Web evidence retrieved by the agent.}
\end{table*}

\begin{figure*}[!htbp]
\centering
\begin{subfigure}[t]{\textwidth}
  \centering
  \includegraphics[width=0.92\textwidth]{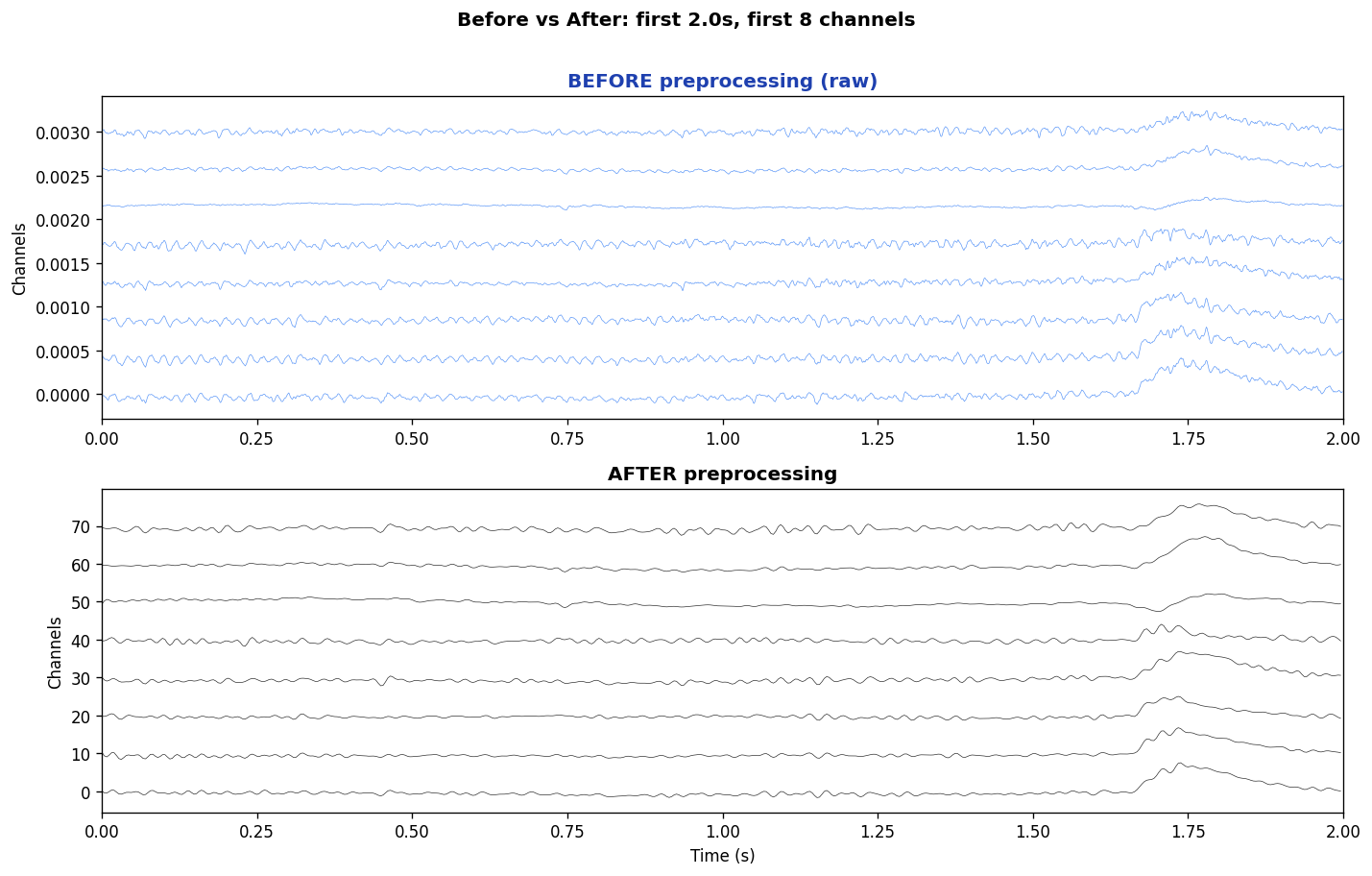}
  \caption{Before-versus-after time-series for the first 2~s on 8 representative channels. The top panel shows the raw signal with visible baseline drift and high-frequency contamination. The bottom panel shows the preprocessed signal with suppressed noise and preserved neural dynamics.}
\end{subfigure}
\vspace{4pt}
\begin{subfigure}[t]{\textwidth}
  \centering
  \includegraphics[width=0.7\textwidth]{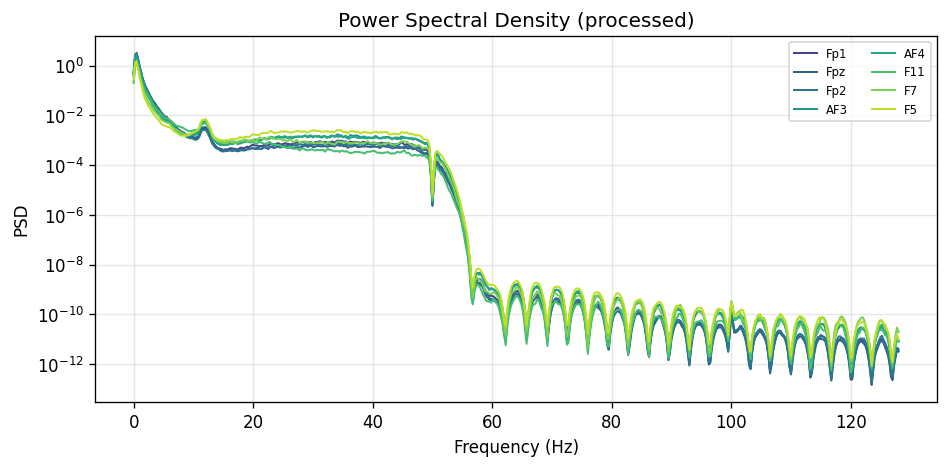}
  \caption{Power spectral density after preprocessing. The 50~Hz peak and harmonics are attenuated by more than four orders of magnitude. Power concentrates in the 0.5 to 45~Hz band with sharp filter roll-off.}
\end{subfigure}
\vspace{4pt}
\begin{subfigure}[t]{\textwidth}
  \centering
  \includegraphics[width=0.92\textwidth]{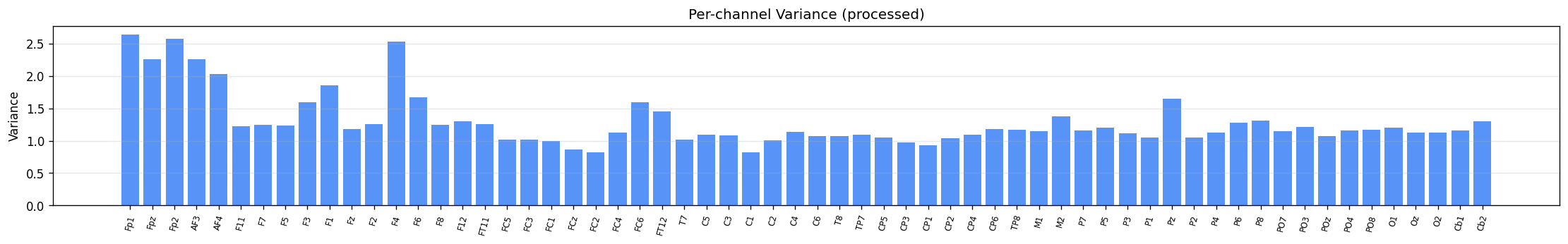}
  \caption{Per-channel variance after preprocessing. Frontal channels (Fp1, Fpz, Fp2, AF3, AF4) retain higher residual variance, consistent with the bad-channel detection reported in Table~\ref{tab:plan}.}
\end{subfigure}
\caption{\textbf{Diagnostic visualisations generated autonomously by EasyBCI on a 64-channel motor imagery recording (Session~1).} All plots are produced by the Reflect stage without human input. \textbf{(A)}~Time-domain comparison shows drift and noise suppression. \textbf{(B)}~Frequency-domain view shows notch filtering and bandpass boundaries. \textbf{(C)}~Spatial variance profile is consistent with channel-level artifact decisions.}
\label{fig:demo-visual}
\end{figure*}

\paragraph{Signal-level diagnostics.}
Figure~\ref{fig:demo-visual} shows the diagnostic visualisations generated by the Reflect stage after pipeline execution. The before-versus-after time-series comparison (Fig.~\ref{fig:demo-visual}A) shows that slow baseline drift and high-frequency contamination visible in the raw trace are suppressed after the nine-step pipeline, while event-related waveform structure is preserved. The post-processing power spectral density (Fig.~\ref{fig:demo-visual}B) shows the notch filter response. The 50~Hz line-noise peak and its harmonics at 100, 150 and 200~Hz are attenuated by more than four orders of magnitude relative to the raw spectrum. Spectral power concentrates in the 0.5 to 45~Hz analysis band with a clean roll-off at the filter boundary. The overall PSD signal-to-noise ratio improves approximately $9\times$ (Table~\ref{tab:qc}). The per-channel variance topography (Fig.~\ref{fig:demo-visual}C) corroborates the bad-channel detection made during planning.

Table~\ref{tab:qc} reports the automatic QC metrics. Across both sessions the pipeline reduces channel count from 67 to 59 (one trigger, five noisy frontal electrodes and two EOG channels removed after ICA), compresses the sampling rate from 1000 to 256~Hz, and improves the PSD signal-to-noise ratio by approximately $9\times$ (from 0.017 to 0.154). No NaN values are introduced, indicating numerical stability through all nine operators.

\begin{table}[!htbp]
\centering
\caption{\textbf{Automatic QC metrics before and after preprocessing.} Both sessions pass all quality checks. PSD SNR improves approximately $9\times$ and no numerical artefacts are introduced.}
\label{tab:qc}
\footnotesize
\setlength{\tabcolsep}{4pt}
\begin{tabular}{@{}l cc cc@{}}
\toprule
& \multicolumn{2}{c}{\textbf{Session 1}} & \multicolumn{2}{c}{\textbf{Session 2}} \\
\cmidrule(lr){2-3} \cmidrule(lr){4-5}
\textbf{Metric} & Before & After & Before & After \\
\midrule
Channels & 67 & 59 & 67 & 59 \\
Sampling rate (Hz) & 1000 & 256 & 1000 & 256 \\
Samples & 2{,}259{,}100 & 578{,}330 & 1{,}459{,}400 & 373{,}606 \\
PSD SNR & 0.017 & 0.154 & 0.018 & 0.155 \\
NaN fraction & 0\% & 0\% & 0\% & 0\% \\
\midrule
\textbf{Verdict} & \multicolumn{2}{c}{\good{PASS}} & \multicolumn{2}{c}{\good{PASS}} \\
\bottomrule
\end{tabular}
\end{table}

\paragraph{Output structure and reproducibility.}
The Execution Agent emits a self-contained mini-repository (Fig.~\ref{fig:code-structure}) satisfying the output contract defined in Section~\ref{sec:method-contract}. Three constraints are enforced at generation time. No import from the EasyBCI framework is permitted (Code Standard Rule~15), so the generated code runs on a bare installation of \texttt{mne}, \texttt{numpy}, \texttt{scipy}, \texttt{scikit-learn} and \texttt{matplotlib}. A deterministic seed (\texttt{EASYBCI\_SEED = 42}) precedes all stochastic operations. The Source Data Guard (Section~\ref{sec:method-guard}) verifies source-file integrity around each step. The exported mini-repository additionally records a SHA-256 hash of the source in \texttt{plan/input\_ref.json} for post-hoc reproducibility.

\paragraph{Scientific reasoning trace.}
Beyond executable code, EasyBCI produces a reasoning document (\texttt{plan/reasoning.md}) that records the scientific rationale behind every preprocessing decision. For each operator in the pipeline, the document states the signal-level motivation for the step, the parameter rationale specific to this recording, and the supporting evidence. Figure~\ref{fig:reasoning-excerpt} shows representative excerpts from the EEG case. A domain expert can audit the rationale behind each decision and revise individual parameter choices.

\begin{figure}[!htbp]
\centering
\begin{tcolorbox}[enhanced, colback=CaseGray, colframe=DeepBlue!70, boxrule=0.4pt,
  left=2mm, right=2mm, top=1.5mm, bottom=1.5mm, width=\columnwidth,
  title={\small\bfseries plan/reasoning.md (excerpt, verbatim agent output)}]
\small
\textbf{Step 2. Notch Filter (Power Line Removal)}

Power spectral density analysis reveals a strong 50~Hz power line peak at 45.8~dB above the noise floor with harmonics at 100, 150, and 200~Hz. This level of line noise contamination would dominate the signal and degrade classification performance. A notch filter at 50~Hz is the standard approach to suppress mains interference while preserving the neural signal in adjacent frequency bands.

\vspace{6pt}
\textbf{Step 6. ICA Artifact Removal (EOG, ECG)}

The VEOG and HEOG channels show variance many times the montage median, consistent with strong eye movement and blink artifacts. The inspection also detected a high rate of blink-like detector events in Session~2. ICA decomposition with automatic EOG and ECG component rejection will identify and remove ocular and cardiac artifacts from the EEG channels. This is critical for motor imagery classification where artifactual components can be mistaken for neural activity.

\vspace{6pt}
\textbf{Step 8. Resample (Downsampling)}

The original 1000~Hz sampling rate is oversampled for motor imagery analysis where the relevant frequency content is below 45~Hz. Downsampling to 256~Hz reduces data size by approximately $4\times$, saving memory and computation time for downstream processing while maintaining more than $5\times$ the Nyquist frequency for the highest frequency of interest.
\end{tcolorbox}
\caption{\textbf{Excerpt from the agent-generated reasoning trace (64-channel motor imagery EEG).} Each step documents the signal evidence that motivated the decision, the scientific rationale for the chosen parameters, and the relationship to downstream classification. The document covers all nine pipeline steps.}
\label{fig:reasoning-excerpt}
\end{figure}

\begin{figure}[!htbp]
\centering
\begin{tcolorbox}[enhanced, colback=CaseGray, colframe=DeepBlue!70, boxrule=0.4pt,
  left=2mm, right=2mm, top=1.5mm, bottom=1.5mm, width=\columnwidth,
  title={\small\bfseries EEG\_preprocess\_work\_dir/}]
\ttfamily\scriptsize
\begin{tabular}{@{}ll@{}}
\textcolor{DeepBlue}{code/} & \\
\quad pipeline.py & \textcolor{gray}{9-step preprocessing (standalone)} \\
\quad build\_ai\_ready.py & \textcolor{gray}{epoch segmentation from behavioural CSV} \\
\quad qc.py & \textcolor{gray}{before/after QC metrics and report} \\
\quad vis.py & \textcolor{gray}{5 diagnostic figures per session} \\
\quad run.py & \textcolor{gray}{one-command orchestrator} \\
\quad requirements.txt & \textcolor{gray}{mne, numpy, scipy, sklearn, matplotlib} \\[3pt]
\textcolor{DeepBlue}{plan/} & \\
\quad proposal.json & \textcolor{gray}{pipeline plan with per-parameter evidence} \\
\quad web\_evidence.json & \textcolor{gray}{literature citations with full snippets} \\
\quad reasoning.md & \textcolor{gray}{step-by-step scientific rationale} \\
\quad config.yaml & \textcolor{gray}{reproducible YAML configuration} \\
\quad input\_ref.json & \textcolor{gray}{SHA-256 hash of source file} \\[3pt]
\textcolor{DeepBlue}{preprocessed\_output/} & \\
\quad preprocessed/ & \textcolor{gray}{continuous NWB files (BIDS layout)} \\
\quad AI\_ready/ & \textcolor{gray}{epoched pickle arrays per condition} \\
\quad QC\_out/ & \textcolor{gray}{JSON and Markdown QC reports} \\
\quad figures/ & \textcolor{gray}{PSD, variance and timeseries plots} \\
\end{tabular}
\end{tcolorbox}
\caption{\textbf{Output mini-repository.} The agent emits a self-contained work directory satisfying the contract in Section~\ref{sec:method-contract}. The \texttt{code/} subtree runs independently of EasyBCI. The \texttt{plan/} subtree preserves the full decision provenance chain. The \texttt{preprocessed\_output/} subtree delivers AI-ready data in NWB and pickle format with quality reports and diagnostic figures.}
\label{fig:code-structure}
\end{figure}

\subsection{Multi-modal Case Studies}
\label{sec:multimodal}

Beyond EEG, we run EasyBCI on five additional modalities to evaluate cross-modal generalisation (Table~\ref{tab:case-summary}). All five are evaluated through signal-level QC and domain-appropriate diagnostics. Full pipeline details, planning tables and diagnostic figures for each case appear in Appendix~\ref{sec:appendix-cases}.

\begin{table*}[t]
\centering
\caption{\textbf{Multi-modal case study summary.} EasyBCI produced a QC-passing pipeline for each of the six modalities under a single natural-language instruction and without per-run reconfiguration; pipeline length varied from 4 to 9 steps.}
\label{tab:case-summary}
\footnotesize
\setlength{\tabcolsep}{4pt}
\begin{tabularx}{\textwidth}{@{}c l c c l c c X@{}}
\toprule
\textbf{\#} & \textbf{Modality} & \textbf{Ch.} & \textbf{$f_s$ (kHz)} & \textbf{Goal} & \textbf{Steps} & \textbf{QC} & \textbf{Key validation} \\
\midrule
1 & Scalp EEG (64 ch) & 64 & 1 & Classification & 9 & \good{PASS} & See Table~\ref{tab:main-results} \\
2 & Unit-level spikes (Neuropixel) & 384 & 30 & Generic & 4 & \good{PASS} & Spatial firing-rate structure preserved \\
3 & sEEG (ds003029) & 135 & 1 & Classification$^\ddagger$ & 7 & \good{PASS} & HFO band preserved, bipolar re-referencing \\
4 & MEG$^\dagger$ (ds000246) & 193 & 1 & Classification & 7 & \good{PASS} & M100/M200 latency preserved, SNR $3.8\times$ \\
5 & fNIRS$^\dagger$ (ds004830) & 84 & 0.05 & Classification & 4 & \good{PASS} & HRF timing preserved, SNR $2.4\times$ \\
6 & ECoG$^\dagger$ (speech) & 124 & 0.512 & Classification (speech) & 7 & \good{PASS} & High-$\gamma$ envelope, SNR $11.1\times$ \\
\bottomrule
\end{tabularx}
\vspace{2pt}
{\raggedright\scriptsize $\dagger$ Classification labels available; quantitative LDA evaluation deferred to future work. $\ddagger$ Classification goal but continuous recording with sparse seizure events; trial-level CV not applicable.\par}
\end{table*}

\paragraph{Neuropixel extracellular recording.}

The user issues the following instruction.

\begin{quote}
\small\itshape
Run threshold-based spike detection on the Neuropixel recording in \texttt{./Neuropixel/}. Output per-unit firing rates binned at 50~Hz (20~ms bins).
\end{quote}

\noindent This instruction carries no classification label and no downstream training objective. The agent detects the absence of a goal keyword, sets \texttt{analysis\_goal = generic}, and restricts output to signal conditioning without epoch segmentation. The input is a single-session NWB file containing 384-channel Neuropixel~1.0 recordings sampled at 30~kHz over 20~s.

The Plan Agent identifies the modality as extracellular spike recording based on the 30~kHz sampling rate and 384-channel count. The agent emits a four-step Pipeline Plan consisting of bandpass filtering (300 to 6000~Hz), median CAR, MAD-based threshold spike detection ($k = 5$), and MUA binning at 20~ms. All 384 channels pass QC. The near-zero post-CAR mean ($1.7 \times 10^{-11}$~\textmu V) is consistent with common-mode rejection. Diagnostic figures and planning details appear in Appendix~\ref{sec:appendix-neuropixel}.

\paragraph{sEEG, MEG, fNIRS and ECoG.}

For intracranial sEEG (ds003029, 135 channels at 1~kHz, epilepsy classification), EasyBCI generates a seven-step pipeline consisting of bad-channel removal, non-data channel exclusion, bipolar re-referencing across adjacent contacts, 50~Hz notch filtering, 0.5 to 200~Hz bandpass, resampling to 500~Hz and robust scaling. The bipolar montage is selected based on the sEEG electrode geometry identified from channel naming conventions during fingerprint computation. The pipeline retains the full high-frequency oscillation band (80 to 200~Hz) required for seizure detection while reducing channel count from 135 to 133 and improving PSD SNR from 0.042 to 0.057.

For MEG (ds000246, 193 channels at 1~kHz, auditory ERP classification), EasyBCI produces a seven-step pipeline covering non-data channel removal, 60~Hz notch filtering (detected from the JSON sidecar indicating a US acquisition site), 1 to 40~Hz bandpass, bad-channel detection, ICA-based EOG and ECG artifact removal, resampling to 250~Hz and robust scaling. The pipeline reduces channel count from 193 to 156, compresses sampling rate by $4\times$, and improves PSD SNR from 0.071 to 0.268, a $3.8\times$ improvement reflecting suppression of line noise and physiological artifacts while retaining auditory evoked components.

For fNIRS (ds004830, 84 channels at 50~Hz, cognitive classification), EasyBCI generates a four-step pipeline. It converts raw intensity to optical density, applies the modified Beer-Lambert law with wavelength-specific parameters (690 and 830~nm) detected from BIDS metadata, applies a 0.01 to 0.5~Hz IIR bandpass to isolate the haemodynamic response while attenuating cardiac oscillations above 0.5~Hz, and performs standard scaling. The pipeline reduces from 84 to 77 channels and improves PSD SNR from 0.071 to 0.171 ($2.4\times$), consistent with suppression of high-frequency noise while retaining haemodynamic response timing.

For ECoG (speech decoding, 124 channels at 512~Hz), EasyBCI generates a seven-step pipeline for high-gamma envelope extraction. The pipeline removes non-data channels (12 DC and 1 ECG), detects and excludes one bad channel by variance criteria (G34 at $11.3\times$ median), applies common average reference, extracts the 70 to 200~Hz band, computes the Hilbert amplitude envelope, resamples from 512 to 100~Hz, and applies robust scaling. This pipeline produces the highest SNR improvement among all modalities ($11.1\times$, from 0.062 to 0.692), consistent with narrowband extraction concentrating energy in the 70 to 200~Hz target band.

Across the six modalities, the same framework produced a QC-passing pipeline without per-run manual reconfiguration. The inputs span nearly three orders of magnitude in sampling rate (50~Hz fNIRS to 30~kHz Neuropixel), channel counts from 64 to 384, and different signal physics (electromagnetic, haemodynamic, extracellular). Pipeline length ranged from 4 steps (unit-level spike generic conditioning) to 9 steps (EEG motor imagery classification).

\section{Conclusion and Discussion}

This work introduced EasyBCI, a two-phase agent that plans and executes preprocessing pipelines for six neural signal modalities under large language model guidance. Under a fixed-classifier protocol that isolates preprocessing as the sole variable, all eight EasyBCI backbones outperform the Manual pipeline on the 4-class task, and six of eight do so on binary classification. Controlled same-backbone comparisons (EasyBCI vs.\ Claude Code on Opus 4.8; EasyBCI vs.\ Codex on GPT-5.5) indicate that domain-specific orchestration preserves more task-relevant linear separability than general-purpose agents given the same model. Among the three sub-40B backbones, Agents-A1 scores highest on both tasks and exceeds Claude Code on the 4-class task. The architecture advantage is consistent across all five frontier-scale backbones but does not extend to all sub-40B backbones. For the preprocessing tasks studied here, the results suggest that among frontier-scale backbones, domain constraints, planning mechanisms and quality feedback contribute more to preprocessing quality than differences in backbone model capability alone. The architectural mechanisms (modality-aware routing, raw-data isolation, quality-weighted skill retrieval with deprecation, and expert oversight at defined decision points) do not increase the model's generative capability. They constrain the decision space, inject domain knowledge at planning time and accumulate empirical evidence. These principles may apply to other scientific workflows in which preprocessing decisions shape downstream conclusions.

Several directions remain open. First, preprocessing and downstream decoding are coupled in practice. Up to 42\% of trial-level predictions flip when only the preprocessing changes~\cite{hou2026same}, yet EasyBCI currently selects parameters without explicit knowledge of the downstream model. Quality-weighted skill retrieval partially addresses this coupling by retaining pipelines whose quality metrics pass. Incorporating downstream decoder performance as an acceptance criterion would close the loop further. Second, the automatic quality-control metrics employed in this work operate at the signal level. Extending these to correlate preprocessing quality with diagnostic outcomes across patient populations and longitudinal recording conditions would strengthen clinical applicability. Third, real-time closed-loop systems impose latency constraints that the current batch-oriented executor does not address. Adapting the plan-execute-verify cycle to real-time latency constraints while preserving auditability remains an open challenge.

Together these directions point toward systems in which preprocessing agents accumulate validated institutional knowledge and adapt to downstream requirements. Achieving this will require static analysis and test-based verification of agent-generated code, longitudinal tracking of skill-library evolution, and coupling with algorithm-selection agents for end-to-end optimisation. The architectural commitments of raw-data isolation, quality-gated retention and expert oversight at defined decision points provide a foundation for these extensions.

\begingroup
\sloppy
\clearpage
\printbibliography[heading=bibintoc]
\endgroup

\clearpage
\appendix
\appendix

\section{Multi-modal Case Study Details}
\label{sec:appendix-cases}

This appendix presents the full pipeline specifications, QC metrics and selected reasoning excerpts for the five non-EEG modalities summarised in Table~\ref{tab:case-summary}. Each case runs EasyBCI in cold-start mode (empty skill library) with a single natural-language instruction. The agent auto-detects the modality, drafts a pipeline plan, executes it and produces a QC report without manual intervention.

\subsection{Neuropixel Extracellular Recording}
\label{sec:appendix-neuropixel}

\paragraph{Dataset.}
A single-session NWB file containing 384 channels from a Neuropixel~1.0 probe, sampled at 30~kHz over 20~s of extracellular recording.

\paragraph{User instruction.}
\begin{quote}
\small\itshape
Run threshold-based spike detection on the Neuropixel recording in \texttt{./Neuropixel/}. Output per-unit firing rates binned at 50~Hz (20~ms bins).
\end{quote}

\paragraph{Pipeline plan.}
The Plan Agent identifies the modality as extracellular spike recording from the 30~kHz sampling rate and 384-channel count. It sets the analysis goal to generic (no classification target) and produces a four-step plan (Table~\ref{tab:plan-neuropixel}).

\begin{table}[H]
\centering
\caption{\textbf{Agent-generated pipeline for Neuropixel spike detection.}}
\label{tab:plan-neuropixel}
\footnotesize
\setlength{\tabcolsep}{3pt}
\begin{tabularx}{\columnwidth}{@{}c l l X@{}}
\toprule
\textbf{\#} & \textbf{Operator} & \textbf{Parameters} & \textbf{Rationale (abridged)} \\
\midrule
1 & \texttt{bandpass} & 300--6000~Hz & Canonical Neuropixel AP band. Removes LFP drift, preserves spike waveform energy \\
2 & \texttt{car} & -- & Subtracts whole-probe mean to suppress spatially correlated noise \\
3 & \texttt{threshold\_spike} & $k\!=\!5$, negative, per-channel & IBL Brain-Wide-Map default for Neuropixels. Per-channel MAD accounts for region-dependent noise \\
4 & \texttt{mua\_binning} & 20~ms, smoothed, rate & User-requested 50~Hz output. Gaussian smoothing suppresses histogram noise \\
\bottomrule
\end{tabularx}
\end{table}

\paragraph{QC metrics.}

\begin{table}[H]
\centering
\caption{\textbf{QC report for Neuropixel recording.}}
\label{tab:qc-neuropixel}
\footnotesize
\begin{tabular}{@{}l cc@{}}
\toprule
\textbf{Metric} & \textbf{Before} & \textbf{After} \\
\midrule
Channels & 384 & 384 \\
Sampling rate (kHz) & 30 & 30 \\
Samples & 600{,}000 & 600{,}000 \\
Mean ($\mu$V) & $-$13.73 & $1.7 \times 10^{-11}$ \\
Std ($\mu$V) & 18.25 & 5.39 \\
PSD SNR & 0.770 & 0.109 \\
NaN fraction & 0\% & 0\% \\
\midrule
\textbf{Verdict} & \multicolumn{2}{c}{\good{PASS}} \\
\bottomrule
\end{tabular}
\end{table}

The near-zero post-CAR mean ($1.7 \times 10^{-11}$~$\mu$V) is consistent with common-mode rejection across the probe. PSD SNR decreases because spikes are brief transients whose spectral energy distributes broadly across the passband rather than concentrating at discrete oscillatory peaks. In spike-detection pipelines, temporal precision matters more than spectral concentration, so PSD SNR reduction after bandpass filtering is typical. All 384 channels pass QC with no NaN values introduced.

\paragraph{Key reasoning excerpt.}
The agent selects $k=5$ rather than the Quiroga default of $k=4$ because Neuropixel probes span multiple brain regions with heterogeneous noise levels. Per-channel MAD estimation is critical because a global threshold would miss spikes on low-noise channels and over-detect on high-noise channels. The 0.5~ms refractory period blocks spurious double-detections from waveform reflections across adjacent shanks.

\begin{figure}[H]
\centering
\begin{subfigure}[t]{\columnwidth}
  \centering
  \includegraphics[width=\columnwidth]{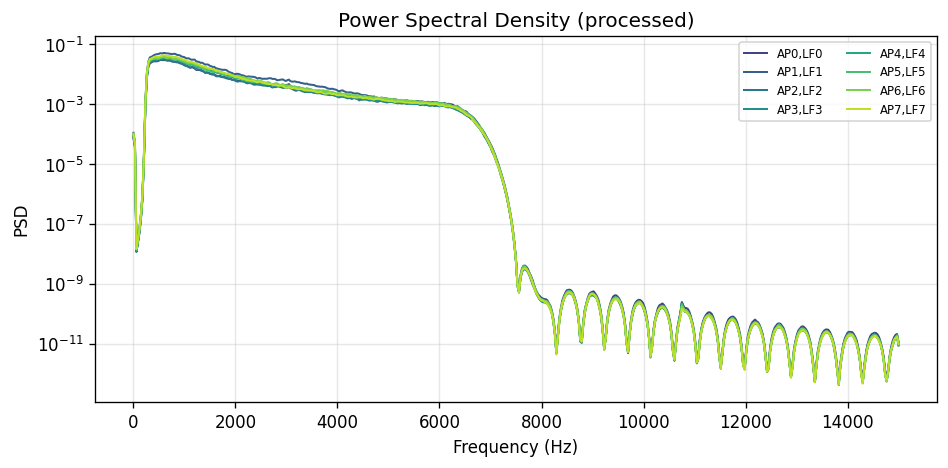}
\end{subfigure}
\vspace{4pt}
\begin{subfigure}[t]{\columnwidth}
  \centering
  \includegraphics[width=\columnwidth]{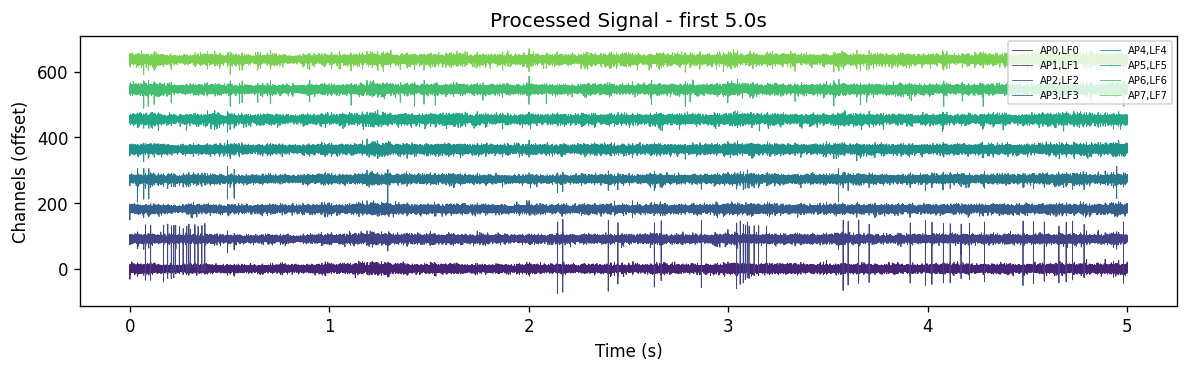}
\end{subfigure}
\caption{\textbf{Neuropixel diagnostic visualisations (384 channels, 30~kHz, 20~s recording).} Generated autonomously by the Reflect stage. \textbf{(A)}~Power spectral density after bandpass filtering (300--6000~Hz) and CAR. Energy concentrates in the action-potential band with steep roll-off at both boundaries. \textbf{(B)}~Processed time-series on 8 channels across the 384-channel probe (first 5~s). Spike transients are visible against the reduced noise floor; firing rates vary across brain regions.}
\label{fig:supp-neuropixel}
\end{figure}

\paragraph{Diagnostic figures.}
Figure~\ref{fig:supp-neuropixel} shows the post-processing diagnostics. The PSD shows spectral energy concentrated in the 300--6000~Hz action-potential band after bandpass filtering, with sharp roll-off at both boundaries. The time-series view shows spike-like transients on representative channels after median CAR, with visible differences in firing rate across brain regions traversed by the probe.

\subsection{sEEG (Stereotactic EEG)}
\label{sec:appendix-seeg}

\paragraph{Dataset.}
OpenNeuro ds003029, subject jh101, presurgical session. A 135-channel intracranial recording at 1~kHz in BrainVision format, acquired for epilepsy monitoring with seizure classification as the intended downstream task.

\paragraph{User instruction.}
\begin{quote}
\small\itshape
Preprocess the sEEG recording in \texttt{./sEEG/} for seizure classification.
\end{quote}

\paragraph{Pipeline plan.}

\begin{table}[H]
\centering
\caption{\textbf{Agent-generated pipeline for sEEG epilepsy classification.}}
\label{tab:plan-seeg}
\footnotesize
\setlength{\tabcolsep}{3pt}
\begin{tabularx}{\columnwidth}{@{}c l l X@{}}
\toprule
\textbf{\#} & \textbf{Operator} & \textbf{Parameters} & \textbf{Rationale (abridged)} \\
\midrule
1 & \texttt{drop\_bads} & auto & Removes flat and saturated channels detected by variance criteria \\
2 & \texttt{drop\_nondata} & data\_only & Removes EKG channels that carry non-neural signal \\
3 & \texttt{bipolar\_ref} & -- & Standard for sEEG depth electrodes. Enhances local field potentials between adjacent contacts \\
4 & \texttt{notch} & 50~Hz & PSD peak at 26.7~dB above noise floor on the 50~Hz power grid \\
5 & \texttt{bandpass} & 0.5--200~Hz & Preserves full HFO band (80--200~Hz) required for seizure detection \\
6 & \texttt{resample} & 500~Hz & Nyquist-safe for 200~Hz upper bound. Halves computation \\
7 & \texttt{scale} & robust & Handles amplitude differences across brain regions without distortion from ictal bursts \\
\bottomrule
\end{tabularx}
\end{table}

\paragraph{QC metrics.}

\begin{table}[H]
\centering
\caption{\textbf{QC report for sEEG recording.}}
\label{tab:qc-seeg}
\footnotesize
\begin{tabular}{@{}l cc@{}}
\toprule
\textbf{Metric} & \textbf{Before} & \textbf{After} \\
\midrule
Channels & 135 & 133 \\
Sampling rate (Hz) & 1000 & 500 \\
Samples & 153{,}000 & 76{,}500 \\
PSD SNR & 0.042 & 0.057 \\
NaN fraction & 0\% & 0\% \\
\midrule
\textbf{Verdict} & \multicolumn{2}{c}{\good{PASS}} \\
\bottomrule
\end{tabular}
\end{table}

The modest SNR improvement (0.042 to 0.057, $1.4\times$) reflects the nature of intracranial recordings, which have inherently high signal quality and low external noise contamination. Bipolar re-referencing enhances spatial selectivity between adjacent depth contacts, the wide 0.5--200~Hz passband preserves the full HFO band for seizure detection, and two non-neural channels (EKG) are removed.

\paragraph{Key reasoning excerpt.}
The agent selects bipolar re-referencing without explicit instruction because it detects sEEG electrode geometry from the channel naming convention (depth contacts with sequential numbering). Bipolar montage subtracts adjacent contacts on the same electrode shaft, cancelling far-field activity and amplifying local neural generators. Robust scaling is chosen over standard scaling because ictal amplitude bursts would inflate the standard deviation and compress inter-ictal features.

\begin{figure}[H]
\centering
\begin{subfigure}[t]{\columnwidth}
  \centering
  \includegraphics[width=\columnwidth]{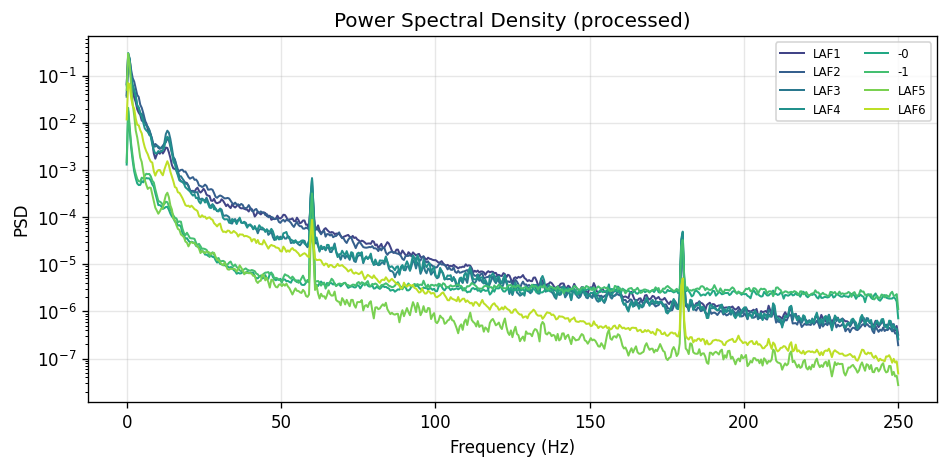}
\end{subfigure}
\vspace{4pt}
\begin{subfigure}[t]{\columnwidth}
  \centering
  \includegraphics[width=\columnwidth]{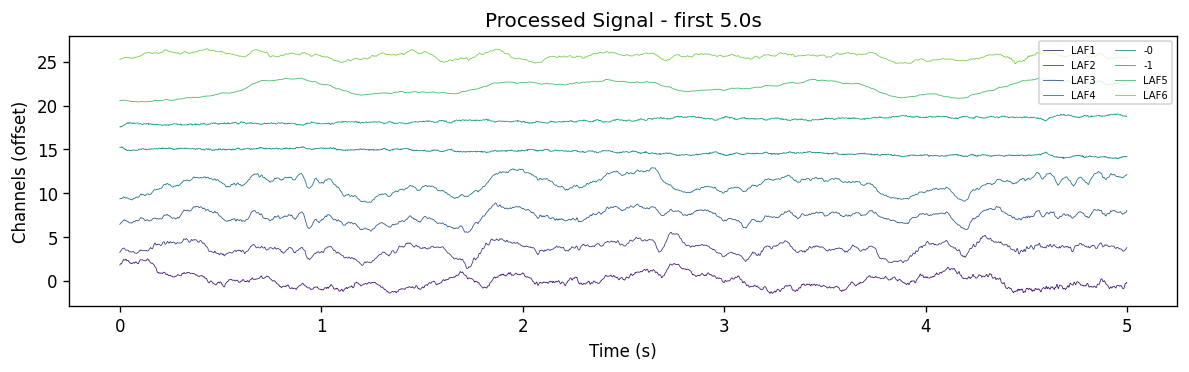}
\end{subfigure}
\caption{\textbf{sEEG diagnostic visualisations (ds003029, subject jh101).} Generated autonomously by the Reflect stage. \textbf{(A)}~Power spectral density after preprocessing. The 50~Hz notch is visible; spectral power spans 0.5--200~Hz, retaining the HFO range for seizure detection. \textbf{(B)}~Processed time-series on 8 bipolar-referenced depth contacts (first 5~s). Stable baselines with no visible drift.}
\label{fig:supp-seeg}
\end{figure}

\paragraph{Diagnostic figures.}
Figure~\ref{fig:supp-seeg} shows the post-processing diagnostics generated by EasyBCI. The PSD shows 50~Hz notch attenuation and preservation of the full 0.5--200~Hz band including the high-frequency oscillation range for seizure detection. The time-series view shows clean bipolar-referenced signals on representative depth contacts with visible neural dynamics across frequency bands.

\subsection{MEG (Magnetoencephalography)}
\label{sec:appendix-meg}

\paragraph{Dataset.}
OpenNeuro ds000246, subject 001. A 193-channel Elekta Neuromag MEG recording at 1~kHz in FIFF format, acquired during an auditory word recognition task. The downstream goal is classification of auditory evoked responses.

\paragraph{User instruction.}
\begin{quote}
\small\itshape
Preprocess the MEG data in \texttt{./MEG/} for auditory ERP classification.
\end{quote}

\paragraph{Pipeline plan.}

\begin{table}[H]
\centering
\caption{\textbf{Agent-generated pipeline for MEG auditory classification.}}
\label{tab:plan-meg}
\footnotesize
\setlength{\tabcolsep}{3pt}
\begin{tabularx}{\columnwidth}{@{}c l l X@{}}
\toprule
\textbf{\#} & \textbf{Operator} & \textbf{Parameters} & \textbf{Rationale (abridged)} \\
\midrule
1 & \texttt{drop\_nondata} & data\_only & Removes 1 STI marker and 35 MISC channels \\
2 & \texttt{notch} & 60~Hz & JSON sidecar reports PowerLineFrequency=60~Hz (US site) \\
3 & \texttt{bandpass} & 1--40~Hz & Preserves M100/M200/N400 components. 1~Hz highpass removes drift before ICA \\
4 & \texttt{drop\_bads} & auto & MEG~056 identified as bad from data metadata \\
5 & \texttt{ica} & eog, ecg & FastICA, 25 components. Removes ocular and cardiac magnetic artifacts \\
6 & \texttt{resample} & 250~Hz & $5\times$ oversampling of 40~Hz upper bound. $4\times$ data reduction \\
7 & \texttt{scale} & robust & Median/IQR normalisation resistant to outlier trials \\
\bottomrule
\end{tabularx}
\end{table}

\paragraph{QC metrics.}

\begin{table}[H]
\centering
\caption{\textbf{QC report for MEG recording.}}
\label{tab:qc-meg}
\footnotesize
\begin{tabular}{@{}l cc@{}}
\toprule
\textbf{Metric} & \textbf{Before} & \textbf{After} \\
\midrule
Channels & 193 & 156 \\
Sampling rate (Hz) & 1000 & 250 \\
Samples & 400{,}000 & 100{,}000 \\
PSD SNR & 0.071 & 0.268 \\
NaN fraction & 0\% & 0\% \\
\midrule
\textbf{Verdict} & \multicolumn{2}{c}{\good{PASS}} \\
\bottomrule
\end{tabular}
\end{table}

The $3.8\times$ SNR improvement results from suppression of 60~Hz line noise, high-frequency environmental noise above 40~Hz, and physiological artifacts (cardiac and ocular) through ICA decomposition. Channel reduction from 193 to 156 results from removing 36 non-data channels (STI and MISC) and 1 bad MEG sensor, retaining all clean magnetometer and gradiometer channels.

\paragraph{Key reasoning excerpt.}
The agent detects the US acquisition site from the JSON sidecar metadata (PowerLineFrequency=60) and sets the notch filter to 60~Hz accordingly. It places the bandpass before ICA because drift removal improves ICA convergence~\cite{gramfort2013meg}. It selects 25 ICA components, the default cap of $\min(n_{\text{channels}}-1, 25)$, for artifact separation from 156 clean MEG channels.

\begin{figure}[H]
\centering
\begin{subfigure}[t]{\columnwidth}
  \centering
  \includegraphics[width=\columnwidth]{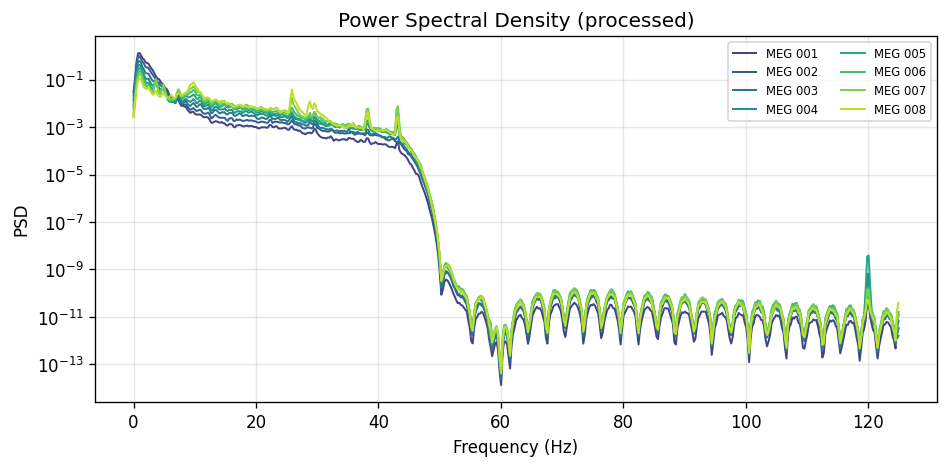}
\end{subfigure}
\vspace{4pt}
\begin{subfigure}[t]{\columnwidth}
  \centering
  \includegraphics[width=\columnwidth]{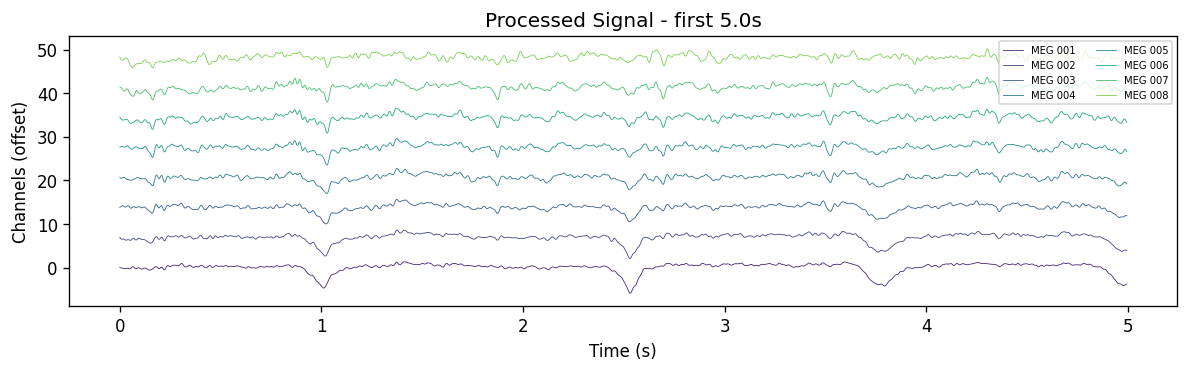}
\end{subfigure}
\caption{\textbf{MEG diagnostic visualisations (ds000246, subject 001, auditory task).} Generated autonomously by the Reflect stage. \textbf{(A)}~Power spectral density after preprocessing. The 60~Hz line-noise peak is attenuated; power concentrates in the 1--40~Hz analysis band. \textbf{(B)}~Processed time-series on 8 magnetometer channels (first 5~s). Stable baselines and preserved evoked-response structure after ICA artifact removal.}
\label{fig:supp-meg}
\end{figure}

\paragraph{Diagnostic figures.}
Figure~\ref{fig:supp-meg} shows the post-processing diagnostics. The PSD shows 60~Hz notch attenuation and clean 1--40~Hz bandpass with sharp roll-off. The time-series view shows stable baselines on representative magnetometer channels after ICA artifact removal.

\subsection{fNIRS (Functional Near-Infrared Spectroscopy)}
\label{sec:appendix-fnirs}

\paragraph{Dataset.}
OpenNeuro ds004830, subject 08. An 84-channel fNIRS recording at 50~Hz in SNIRF format, with dual wavelengths (690 and 830~nm) for haemodynamic imaging during a cognitive task. The downstream goal is classification of cognitive states.

\paragraph{User instruction.}
\begin{quote}
\small\itshape
Preprocess the fNIRS data in \texttt{./fNIRS/} for classification.
\end{quote}

\paragraph{Pipeline plan.}

\begin{table}[H]
\centering
\caption{\textbf{Agent-generated pipeline for fNIRS cognitive classification.}}
\label{tab:plan-fnirs}
\footnotesize
\setlength{\tabcolsep}{3pt}
\begin{tabularx}{\columnwidth}{@{}c l l X@{}}
\toprule
\textbf{\#} & \textbf{Operator} & \textbf{Parameters} & \textbf{Rationale (abridged)} \\
\midrule
1 & \texttt{optical\_density} & baseline=10~s & First 10~s assumed stable. Converts raw intensity to $\Delta$OD \\
2 & \texttt{beer\_lambert} & $d$=3~cm, DPF=6.0, $\lambda$=[690,830] & Standard adult parameters. Wavelengths from BIDS metadata \\
3 & \texttt{bandpass} & 0.01--0.5~Hz, IIR order~4 & Isolates HRF. Attenuates cardiac oscillations above 0.5~Hz \\
4 & \texttt{scale} & standard & Zero-mean, unit-variance for ML input \\
\bottomrule
\end{tabularx}
\end{table}

\paragraph{QC metrics.}

\begin{table}[H]
\centering
\caption{\textbf{QC report for fNIRS recording.}}
\label{tab:qc-fnirs}
\footnotesize
\begin{tabular}{@{}l cc@{}}
\toprule
\textbf{Metric} & \textbf{Before} & \textbf{After} \\
\midrule
Channels & 84 & 77 \\
Sampling rate (Hz) & 50 & 50 \\
Samples & 148{,}450 & 148{,}450 \\
PSD SNR & 0.071 & 0.171 \\
NaN fraction & 0\% & 0\% \\
\midrule
\textbf{Verdict} & \multicolumn{2}{c}{\good{PASS}} \\
\bottomrule
\end{tabular}
\end{table}

The $2.4\times$ SNR improvement results from suppression of cardiac pulsatility (above 0.5~Hz) and slow instrumental drift (below 0.01~Hz), concentrating spectral energy in the haemodynamic response band. Seven channels are dropped by the optical-density stage's built-in quality gate (channels whose raw intensity is too low or too saturated for reliable Beer-Lambert inversion). The sampling rate is preserved at 50~Hz because the haemodynamic response already operates at this timescale and further downsampling would attenuate the signal of interest.

\paragraph{Key reasoning excerpt.}
The agent detects wavelengths directly from BIDS \texttt{channels.tsv} metadata and applies wavelength-specific parameters for the Beer-Lambert conversion. It selects IIR (4th-order Butterworth) over FIR filtering because the 50~Hz sampling rate would require impractically long FIR filter lengths to achieve adequate frequency resolution at the 0.01~Hz lower cutoff. The 0.5~Hz upper cutoff is chosen to attenuate cardiac oscillations (typically 1--1.5~Hz) while preserving the haemodynamic response function whose power concentrates below 0.2~Hz.

\begin{figure}[H]
\centering
\begin{subfigure}[t]{\columnwidth}
  \centering
  \includegraphics[width=\columnwidth]{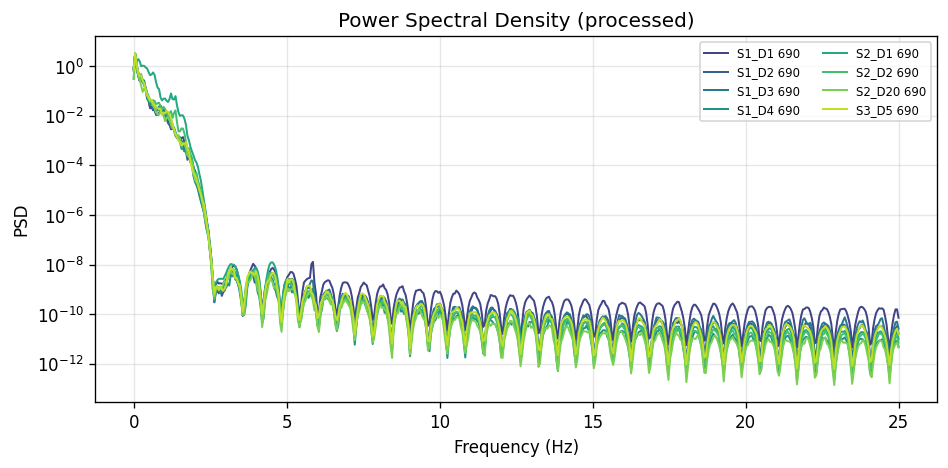}
\end{subfigure}
\vspace{4pt}
\begin{subfigure}[t]{\columnwidth}
  \centering
  \includegraphics[width=\columnwidth]{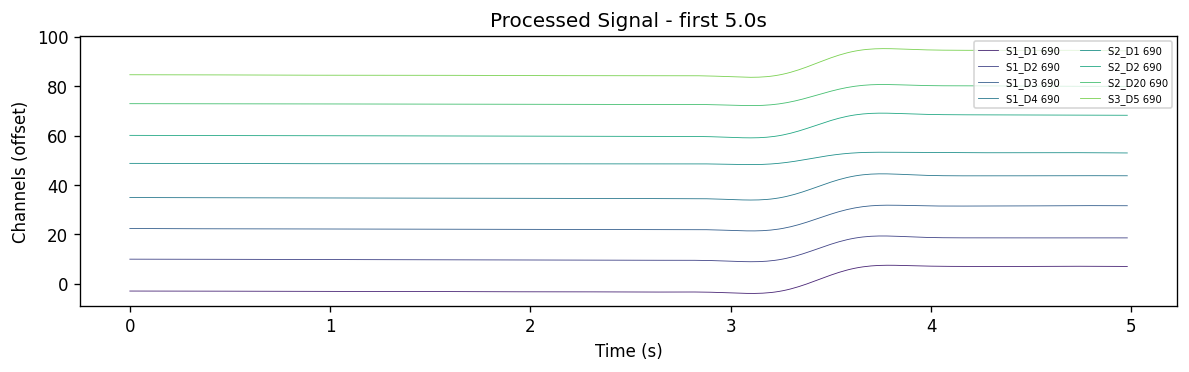}
\end{subfigure}
\caption{\textbf{fNIRS diagnostic visualisations (ds004830, subject 08, cognitive task).} Generated autonomously by the Reflect stage. \textbf{(A)}~Power spectral density after preprocessing. Spectral energy concentrates below 0.5~Hz (haemodynamic band); cardiac oscillations above 1~Hz are attenuated. \textbf{(B)}~Processed time-series on 8 fNIRS channels (690~nm, first 5~s). Smooth, drift-free baselines with visible haemodynamic responses.}
\label{fig:supp-fnirs}
\end{figure}

\paragraph{Diagnostic figures.}
Figure~\ref{fig:supp-fnirs} shows the post-processing diagnostics. The PSD shows spectral power concentrated below the 0.5~Hz upper cutoff with cardiac pulsatility suppressed. The time-series view shows the slow haemodynamic response waveform on representative channels after Beer-Lambert conversion and bandpass filtering.

\subsection{ECoG (Electrocorticography)}
\label{sec:appendix-ecog}

\paragraph{Dataset.}
A 124-channel subdural ECoG recording at 512~Hz in EDF format from a speech decoding study. Electrodes include a 64-contact grid (G1--G64), cortical strips (AT, MT, PT, OC, FR) and depth electrodes (DAMT, DPMT), plus 12 DC channels and 2 ECG channels. The downstream goal is high-gamma feature extraction for speech decoding.

\paragraph{User instruction.}
\begin{quote}
\small\itshape
Preprocess the ECoG recording in \texttt{./ECoG/} for speech decoding. Extract high-gamma features.
\end{quote}

\paragraph{Pipeline plan.}

\begin{table}[H]
\centering
\caption{\textbf{Agent-generated pipeline for ECoG speech decoding.}}
\label{tab:plan-ecog}
\footnotesize
\setlength{\tabcolsep}{3pt}
\begin{tabularx}{\columnwidth}{@{}c l l X@{}}
\toprule
\textbf{\#} & \textbf{Operator} & \textbf{Parameters} & \textbf{Rationale (abridged)} \\
\midrule
1 & \texttt{drop\_nondata} & data\_only & Removes 12 DC (MISC) and 1 ECG channel \\
2 & \texttt{drop\_bads} & auto & G34 at $11.3\times$ median variance \\
3 & \texttt{car} & -- & Removes shared reference noise across 110 clean ECoG channels \\
4 & \texttt{bandpass} & 70--200~Hz & Canonical high-$\gamma$ band for speech decoding \\
5 & \texttt{hilbert} & envelope & Extracts amplitude envelope. Correlates with population firing rate \\
6 & \texttt{resample} & 100~Hz & Envelope bandwidth $<$50~Hz. Standard for speech BCI \\
7 & \texttt{scale} & robust & Handles electrode impedance variation and residual outliers \\
\bottomrule
\end{tabularx}
\end{table}

\paragraph{QC metrics.}

\begin{table}[H]
\centering
\caption{\textbf{QC report for ECoG recording.}}
\label{tab:qc-ecog}
\footnotesize
\begin{tabular}{@{}l cc@{}}
\toprule
\textbf{Metric} & \textbf{Before} & \textbf{After} \\
\midrule
Channels & 124 & 110 \\
Sampling rate (Hz) & 512 & 100 \\
Samples & 921{,}600 & 180{,}000 \\
PSD SNR & 0.062 & 0.692 \\
NaN fraction & 0\% & 0\% \\
\midrule
\textbf{Verdict} & \multicolumn{2}{c}{\good{PASS}} \\
\bottomrule
\end{tabular}
\end{table}

The $11.1\times$ SNR improvement is the largest among all modalities, resulting from narrowband high-gamma extraction from broadband ECoG. The 70--200~Hz bandpass removes all low-frequency oscillatory power and drift, while the Hilbert envelope conversion concentrates energy in a slowly varying amplitude signal whose spectral content is below 50~Hz. Channel reduction from 124 to 110 removes 13 non-neural channels (12 DC and 1 ECG) in step~1 and 1 bad ECoG channel (G34, std=$407.6~\mu$V versus a montage median of $36~\mu$V) in step~2.

\paragraph{Key reasoning excerpt.}
The agent selects the 70--200~Hz band based on the speech-decoding goal, citing published evidence that high-gamma amplitude is the primary proxy for population firing rate in speech BCI. The 512~Hz sampling rate provides adequate Nyquist margin (256~Hz) for the 200~Hz upper cutoff. Resampling to 100~Hz after envelope extraction is justified by the observation that speech features (phonemes, words) evolve at 2--20~Hz, making the envelope bandwidth well below 50~Hz.

\begin{figure}[H]
\centering
\begin{subfigure}[t]{\columnwidth}
  \centering
  \includegraphics[width=\columnwidth]{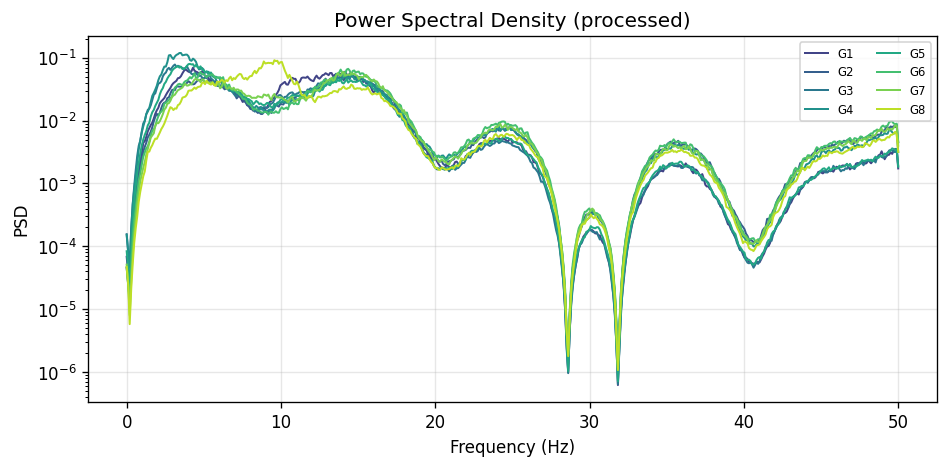}
\end{subfigure}
\vspace{4pt}
\begin{subfigure}[t]{\columnwidth}
  \centering
  \includegraphics[width=\columnwidth]{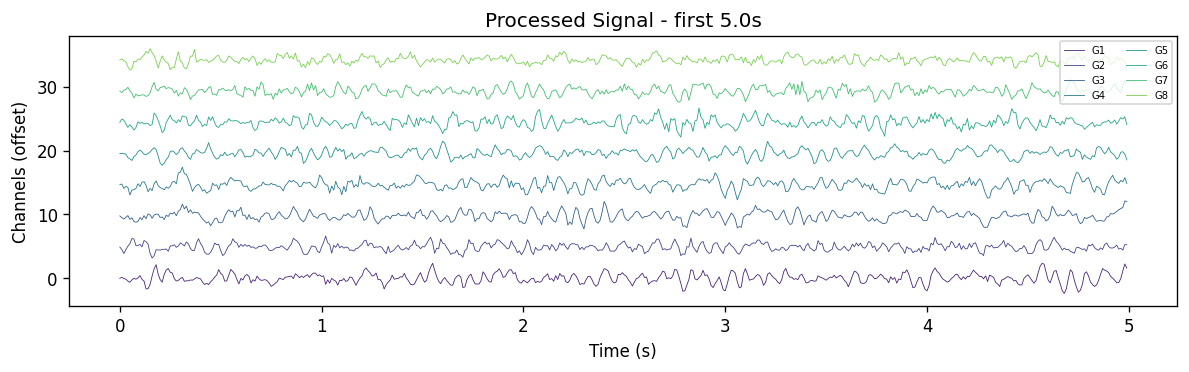}
\end{subfigure}
\vspace{4pt}
\begin{subfigure}[t]{\columnwidth}
  \centering
  \includegraphics[width=\columnwidth]{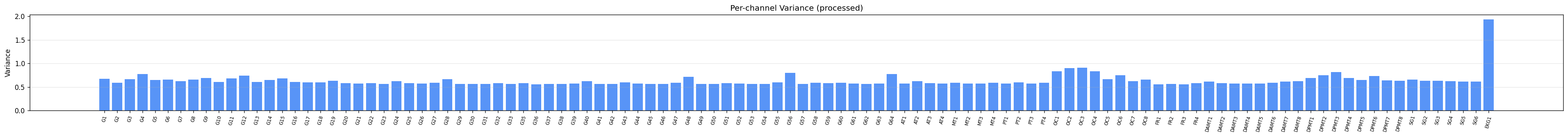}
\end{subfigure}
\caption{\textbf{ECoG diagnostic visualisations (speech decoding, 124 channels).} Generated autonomously by the Reflect stage. The pipeline extracts high-gamma amplitude envelope (70--200~Hz) and resamples to 100~Hz for speech BCI feature extraction. \textbf{(A)}~Power spectral density of the high-gamma envelope (Hilbert extraction, resampled to 100~Hz). Energy concentrates below 50~Hz, matching speech articulation dynamics. \textbf{(B)}~High-gamma envelope on 8 grid electrodes (G1--G8, first 5~s). Smooth, slowly varying signals suitable for speech decoding. \textbf{(C)}~Per-channel variance after preprocessing. Most channels cluster near the median; elevated variance on rightmost channels corresponds to strip electrodes at tissue boundaries.}
\label{fig:supp-ecog}
\end{figure}

\paragraph{Diagnostic figures.}
Figure~\ref{fig:supp-ecog} shows the post-processing diagnostics. The PSD shows the high-gamma envelope's spectral concentration below 50~Hz after Hilbert extraction, consistent with isolation of speech-relevant amplitude modulation. The time-series view shows the smooth, slowly varying high-gamma envelope on representative grid electrodes. The per-channel variance profile shows the distribution across the 110 retained channels after bad-channel exclusion.

\end{document}